%Paper: hep-th/9401074
%From: BREKKE@UICWS.PHY.UIC.EDU
%Date: Fri, 14 Jan 1994 16:55:05 -0600 (CST)

\input harvmac

% Poor man's Blackboard Bold characters often used :
\def\inbar{\,\vrule height1.5ex width.4pt depth0pt}
\def\IB{\relax{\rm I\kern-.18em B}}
\def\IC{\relax\hbox{$\inbar\kern-.3em{\rm C}$}}
\def\ID{\relax{\rm I\kern-.18em D}}
\def\IE{\relax{\rm I\kern-.18em E}}
\def\IF{\relax{\rm I\kern-.18em F}}
\def\IG{\relax\hbox{$\inbar\kern-.3em{\rm G}$}}
\def\IH{\relax{\rm I\kern-.18em H}}
\def\II{\relax{\rm I\kern-.18em I}}
\def\IK{\relax{\rm I\kern-.18em K}}
\def\IL{\relax{\rm I\kern-.18em L}}
\def\IM{\relax{\rm I\kern-.18em M}}
\def\IN{\relax{\rm I\kern-.18em N}}
\def\IO{\relax\hbox{$\inbar\kern-.3em{\rm O}$}}
\def\IP{\relax{\rm I\kern-.18em P}}
\def\IQ{\relax\hbox{$\inbar\kern-.3em{\rm Q}$}}
\def\IR{\relax{\rm I\kern-.18em R}}
\font\cmss=cmss10 \font\cmsss=cmss10 at 7pt
\def\IZ{\relax\ifmmode\mathchoice
{\hbox{\cmss Z\kern-.4em Z}}{\hbox{\cmss Z\kern-.4em Z}}
{\lower.9pt\hbox{\cmsss Z\kern-.4em Z}}
{\lower1.2pt\hbox{\cmsss Z\kern-.4em Z}}\else{\cmss Z\kern-.4em Z}\fi}
\def\IGa{\relax\hbox{${\rm I}\kern-.18em\Gamma$}}
\def\IPi{\relax\hbox{${\rm I}\kern-.18em\Pi$}}
\def\ITh{\relax\hbox{$\inbar\kern-.3em\Theta$}}
\def\IOm{\relax\hbox{$\inbar\kern-3.00pt\Omega$}}

\font\cmss=cmss10

\Title{\vbox{\baselineskip12pt\hbox{HUTP-93/A037; UICHEP-TH/93-18;
BUHEP-93-29}}}
{\vbox{\centerline{Spinning Particles, Braid Groups and Solitons}}}
%   \footnote{}{*optional footnote on title}

\centerline{Lee Brekke$^a$}
\vskip .2 in
\centerline{Michael J. Dugan$^{b,c}$}
\centerline{and}
\centerline{Tom D. Imbo$^{a}$}

\footnote{}{$^a$Department of Physics, University of Illinois at Chicago,
Chicago, IL \ 60607-7059}

\footnote{}{$^b$Lyman Laboratory of Physics,
Harvard University, Cambridge, MA \ 02138}

\footnote{}{$^c$Department of Physics, Boston University, 590 Commonwealth
Avenue, Boston, MA \ 02215}

%if too many authors for abstract on same page, say   \vfill\eject\pageno0

\vskip .5 in

We develop general techniques for computing the fundamental group of the
configuration space of $n$ identical particles, possessing a generic internal
structure, moving on a manifold $M$. This group generalizes the $n$-string
braid group of $M$ which is the relevant object for structureless particles.
In particular, we compute these generalized braid groups for particles with an
internal spin degree of freedom on an arbitrary $M$. A study of their
unitary representations allows us to determine the available spectrum of spin
and statistics on $M$ in a certain class of quantum theories. One interesting
result is that half-integral spin quantizations are obtained on certain
manifolds having an obstruction to an ordinary spin structure. We also
compare our results to corresponding ones for topological solitons in
$O(d+1)$-invariant nonlinear sigma models in $(d+1)$-dimensions, generalizing
recent studies in two spatial dimensions. Finally, we prove that there exists
a general scalar quantum theory yielding half-integral spin for particles (or
$O(d+1)$ solitons) on a closed, orientable manifold $M$ if and only if
$M$ possesses a ${\rm spin}_c$ structure.

\Date{}

\newsec{Introduction}

Consider a single particle moving on a smooth manifold $M$, dim $M\geq 2$, and
suppose that this particle has an internal structure. Denote by $Y$
the space parametrizing this structure. In the simplest case, the
configuration space of this one-particle system is the cartesian product
$M\times Y$. However, more generally, the particle's internal coordinates may
be correlated with its spatial position. Here, the configuration space can
be the total space $E$ of a fiber bundle with base space $M$ and fiber $Y$.
If we have $n$ such particles which are distinguishable (and noncoinciding)
the configuration space becomes $E^n-\Delta_p$, where
$$
\Delta_p=\{ (e_1,\dots ,e_n)\in E^n\ |\ p(e_i)=p(e_j)\ \ for\ some\ \ i\neq j\}
$$
and $p:E\rightarrow M$ is the projection map of the above bundle. The
corresponding configuration space for $n$ {\it identical} particles is the
orbit space $Q_n^E(M)\equiv (E^n-\Delta_p)/S_n$, where the permutation group
$S_n$ has the obvious action on $E^n-\Delta_p$. It is this space that will
occupy our attention below.

The fundamental group $\pi_1(Q_n^E(M))$ plays an important role in describing
various quantizations of the $n$ particle system.\foot{Without loss of
generality we assume that $E$, and therefore $Q_n^E(M)$, is path-connected.
Hence, different choices of basepoint in $Q_n^E(M)$ lead to isomorphic
fundamental groups and in what follows we will suppress this choice and simply
write $\pi_1(Q_n^E(M))$.} More precisely, to every irreducible unitary
representation (IUR) $\rho$ of $\pi_1(Q_n^E(M))$ there exists a quantization
of this system\foot{For more on the notion of quantization used
below, see \ref\cji{C.~J.~Isham, in {\it Relativity, Groups and Topology II},
edited by B.~S.~De~Witt and R.~Stora (Elsevier, New York, 1984)\semi
A.~P.~Balachandran, G.~Marmo, B.-S.~Skagerstam and A.~Stern, {\it Classical
Topology and Quantum States} (Singapore, World Scientific, 1991).}\ref\diss
{T.~D.~Imbo, Ph.~D. Dissertation, University of Texas at Austin, 1988.} and
references therein.} whose fixed-time state vectors are sections of an
(irreducible) $\IC^N$-bundle $B$ over $Q_n^E(M)$, $N$ = dim $\rho$. $B$ is
equipped with a flat $U(N)$ connection whose holonomy realizes $\rho$. An
alternative way of viewing these state vectors $\Psi$ is as multivalued
functions from $Q_n^E(M)$ to $\IC^N$, such that when the argument of $\Psi$ is
brought around a loop in the homotopy class $[\ell ]\in\pi_1(Q_n^E(M))$ we have
$\Psi\rightarrow\rho ([\ell ])\Psi$. One may certainly consider more general
quantum theories --- for instance, those associated with {\it nonflat} complex
vector bundles over $Q_n^E(M)$. However, until Section 8 we will restrict
ourselves to those described above since many interesting features can already
be seen at this level.
For example, if the particles are structureless --- that is, $Y$ is just a
point --- then $E=M$ and $\pi_1(Q_n^E(M))$ is the $n$-string braid group
$B_n(M)$ of the manifold $M$ \ref\bg{R.~H.~Fox and L.~Neuwirth, Math. Scand.
{\bf 10} (1962) 111\semi J.~S.~Birman, {\it Braids, Links and Mapping Class
Groups} (Princeton University Press, Princeton, 1974)\semi V.~L.~Hansen,
{\it Braids and Coverings} (Cambridge University Press, Cambridge,
1989).}\ref\bgt{E.~Fadell and L.~Neuwirth, Math. Scand. {\bf 10} (1962) 119.}.
Given an IUR $\rho$ of $B_n(M)$, $n\geq 2$, one can determine the statistics of
the $n$ identical particles in the corresponding quantum theory by restricting
$\rho$ to those elements of $B_n(M)$ representing local permutations of the
particles \ref\ti{T.~D.~Imbo, C.~Shah~Imbo and E.~C.~G.~Sudarshan, Phys. Lett.
{\bf B234} (1990) 103.}. If one considers only {\it scalar} quantum theories
(those associated with the one-dimensional IUR's of $B_n(M)$), then it has been
shown that the only allowed statistics for the particles are Bose and Fermi if
dim $M\geq 3$ (both always being possible) \ti . By contrast, for open
2-manifolds (like $\IR^2$) one obtains the full range of fractional (or
$\theta$-) statistics\foot{For a review of fractional statistics, including
references to earlier literature, see \ref\lech{R.~Iengo and K.~Lechner, Phys.
Rep. {\bf 213} (1992) 179.}.} for any $n\geq 2$ \ref\ug{T.~D.~Imbo,
C.~Shah~Imbo, R.~S.~Mahajan and E.~C.~G.~Sudarshan, {\it A User's Guide to
Exotic Statistics}, Center for Particle Theory preprint (1990).}.
On the 2-sphere $S^2$ only a finite ($n$-dependent) subset of statistical
angles $\theta$ are allowed \ref\tw{D.~J.~Thouless and Y.-S.~Wu, Phys. Rev.
{\bf B31} (1985) 1191\semi J.~S.~Dowker, J. Phys. {\bf A18} (1985) 3521.},
while on all other closed surfaces only Bose and Fermi statistics
($\theta=0$ and $\pi$) are available \ti . However, by looking at
{\it nonscalar} quantum theories (higher-dimensional IUR's of $B_n(M)$), one
can regain all the rational values of $\theta /\pi$
for $n$ identical particles on any closed, orientable 2-manifold
\ref\te{T.~Einarsson, Phys. Rev. Lett. {\bf 64} (1990) 1995.}\ref
\jmr{T.~D.~Imbo and J.~March-Russell, Phys. Lett. {\bf B252} (1990) 84.}\ref
\tte{T.~Einarsson, Mod. Phys. Lett. {\bf B5} (1991) 675.}.
Further, nonscalar quantum theories allow for the well-known parastatistics
when $n\geq 3$ (and any $M$), as well as a complex generalization of them
for three or more particles in two dimensions \ti\ref\cgo{G.~A.~Goldin,
R.~Menikoff and D.~H.~Sharp, Phys. Rev. Lett. {\bf 54} (1985) 603\semi
R.~Cappuccio and E.~Guadagnini, Phys. Lett. {\bf B252} (1990) 420\semi
A.~P.~Balachandran, M.~Bordeaux and S.~Jo, Int. J. Mod. Phys. {\bf A5} (1990)
2423\semi N.~Read and G.~Moore, Nucl. Phys. {\bf B360} (1991)
363.}\ref\ecg{E.~C.~G.~Sudarshan, T.~D.~Imbo and T.~R.~Govindarajan, Phys.
Lett. {\bf B213} (1988) 471\semi E.~C.~G.~Sudarshan, T.~D.~Imbo and
C.~Shah~Imbo, Ann. Inst. Henri Poincar\'e {\bf 49} (1988) 387.}. If $M$ is
not simply connected, then there may be even more exotic possibilities such as
{\it ambistatistics} where the superselection rule between bosons and fermions
is effectively broken \ti . There is also a fractional version of
ambistatistics on nonsimply connected 2-manifolds \jmr .

The groups $\pi_1(Q_n^E(M))\equiv B_n^E(M)$, for a generic fiber bundle $E$
over $M$, generalize the braid groups $B_n(M)$. In this paper we will focus
on the case where $Y$ parametrizes an internal spin degree of freedom for
the particles. If $M$ is an orientable manifold, we choose a fixed
orientation for $M$. Then the relevant $E$ is the bundle of oriented,
orthonormal $d$-frames over $M$, $d$ = dim $M$, whose fiber $Y$ is
homeomorphic to the special orthogonal group $SO(d)$
\tte\ref\ts{R.~D.~Tscheuschner, Int. J. Theor. Phys. {\bf 28} (1989)
1269.}\ref\apb{A.~P.~Balachandran, T.~Einarsson, T.~R.~Govindarajan and
R.~Ramachandran, Mod. Phys. Lett. {\bf A6} (1991) 2801.}. It is the principal
bundle associated with the tangent bundle of $M$. For nonorientable spaces,
$E$ is the bundle of {\it all} $d$-frames over $M$ and $Y$ is the full
orthogonal group $O(d)$. In both cases, we denote this bundle by $F(M)$.
Just as above, we can determine the statistics of the particles associated
with an IUR $\rho$ of $B_n^F(M)$ by restricting to the local permutations.
But now we can also ask about their spin. We obtain this information by
looking at how those elements of $B_n^F(M)$ which correspond to
2$\pi$-rotations of the particles' frames are represented by $\rho$. Note
that there is no reason to expect a spin-statistics relation for these
mechanical systems.

In what follows we will completely calculate the groups $B_n^F(M)$, for
$M$ an orientable manifold of dimension $d\geq 3$, in
terms of $\pi_1(M)$ and information about possible obstructions
to a spin structure on $M$. One consequence of our results is that there
exist theories where the particles have half-integral spin even
though the ambient space $M$ is not a spin manifold. The nonspin
manifolds for which this phenomenon occurs do, however, possess generalized
spin
structures.
We also comment on the nonorientable case, as well as the situation in two
dimensions. Next, we discuss the relationship between the groups
$B_n^F(M)$ and the fundamental group of the configuration space of the
$O(d+1)$-invariant nonlinear sigma model with space manifold $M$. These
models possess topological solitons, and the implications of our results for
their spin and statistics are demonstrated. Finally, we return to the
phenomenon of half-integral spin for particles and solitons on nonspin
manifolds and study it from the point of view of more general quantum
theories. The reader may find it useful to read the conclusions of this paper
(Section 9) before proceeding to the main text. Although it uses some technical
language which is only defined later, it may help to keep the organization
and goals of the paper clearer as he or she proceeds.

We close this section with a word concerning references. In many places in
the text we use reasonably well-known results or structures from the
mathematical literature. Instead of providing a reference at each such
occurrence (many of which are marked with italics), we have decided to give a
general mathematical bibliography here. The results and definitions that we
have utilized from various areas of mathematics can be found (and traced
further) using the treatises cited below and references contained therein. For
discrete group theory, see \ref\gto{D.~J.~S.~Robinson, {\it A Course in the
Theory of Groups} (Springer-Verlag, New York, 1982)\semi D.~L.~Johnson,
{\it Presentations of Groups} (Cambridge University Press, Cambridge, 1990).};
for algebraic topology, see \ref\alt{G.~W.~Whitehead, {\it Elements of Homotopy
Theory} (Springer-Verlag, New York, 1978)\semi W.~S.~Massey, {\it A Basic
Course in Algebraic Topology} (Springer-Verlag, New York, 1991).}
\ref\span{E.~Spanier, {\it Algebraic Topology} (Springer-Verlag, Berlin,
1966).}; for bundle theory and characteristic classes, see
\span\ref\bun{N.~Steenrod, {\it The Topology of Fibre Bundles} (Princeton
University Press, Princeton, 1951)\semi D.~Husemoller, {\it Fibre Bundles}
(Springer-Verlag, New York, 1966)\semi J.~W.~Milnor and J.~D.~Stasheff,
{\it Characteristic Classes} (Princeton University Press, Princeton, 1974).};
for spin structures on manifolds, see \ref\sstr {H.~Blaine~Lawson and
M.-L.~Michelsohn, {\it Spin Geometry} (Princeton University Press, Princeton,
1989)\semi P.~B.~Gilkey, {\it The Geometry of Spherical Space Form Groups}
(World Scientific, Singapore, 1989)\semi B.~Grinstein, J.~Preskill and R.~Rohm,
unpublished Caltech preprint.}. We have attempted to provide additional
references at each occurrence in the text of a result which is less familiar.

\newsec{Computing $B_n^E(M)$ - General Remarks}

Before specializing to the case of interest, we make some remarks on the
computation of $B_n^E(M)$ in general. A useful fact is that the fiber bundle
$Y\buildrel i \over\hookrightarrow E\buildrel p
\over\rightarrow M$ gives rise to a similar structure $Y^n\buildrel i
\over\hookrightarrow Q_n^E(M)\buildrel p_n \over\rightarrow Q_n(M)$
for all $n\geq 1$. Here $Q_n(M)\equiv M^n-\Delta_{id_M}$ is the configuration
space of $n$ identical structureless particles on $M$. The projection map
$p_n:Q_n^E(M)\rightarrow Q_n(M)$ is given by
$p_n([e_1,\dots ,e_n])=[p(e_1),\dots ,p(e_n)]$, where $[x_1,\dots ,x_n]$
denotes an unordered $n$-tuple of points in the appropriate space.
The {\it long exact homotopy sequence} of
this fiber bundle yields\foot{Note that $\pi_0(Y)$ does not, in general,
possess a natural group structure. Thus, for the last two maps in this
sequence we mean exact in the sense of pointed sets.}
\eqn\long{\dots\rightarrow\pi_2(Q_n(M))\buildrel \alpha_n \over
\rightarrow\pi_1(Y)^n\buildrel i_* \over \rightarrow
B_n^E(M)\buildrel (p_n)_* \over\rightarrow B_n(M)\buildrel \beta_n \over
\rightarrow\pi_0(Y)^n\rightarrow \{*\}\ .}
In this sequence, the basepoint $y_0\in Y$ determines the basepoints
$i_*(y_0)\in Q_n^E(M)$ and $p_n(i_*(y_0))\in Q_n(M)$. Of course when $n=1$,
this reduces to the long exact homotopy sequence of the bundle $Y\buildrel i
\over\hookrightarrow E\buildrel p\over\rightarrow M$. The map $\alpha_n$ in
\long\ is called the {\it connecting homomorphism} and there exists a simple
relationship between $\alpha_1:\pi_2(M)\to\pi_1(Y)$ and $\alpha_n$,
$n\geq 2$. More precisely, the inclusion of $M^n-\Delta_{id_M}$ into $M^n$
induces a homomorphism\foot{Since the codimension of $\Delta_{id_M}$ in $M^n$
is equal to the dimension $d$ of $M$, $\mu_n$ is an isomorphism if $d\geq 4$
and an epimorphism if $d=3$.} $\mu_n:\pi_2(Q_n(M))\rightarrow\pi_2(M)^n$, and
the map $\alpha_n$ is simply the composition $(\alpha_1)^n\circ\mu_n$.

More compactly, we may exhibit $B_n^E(M)$ as an {\it extension}:
\eqn\ex{\{ e\}\rightarrow\pi_1(Y)^n/Im\ \alpha_n\buildrel
i_* \over \rightarrow
B_n^E(M)\buildrel (p_n)_* \over \rightarrow Im\ (p_n)_*\rightarrow\{ e\}.}
The following information suffices to determine $B_n^E(M)$ from the exact
sequence \ex . First, we must know the groups $G\equiv
i_*(\pi_1(Y)^n/Im\ \alpha_n)$
and $H\equiv Im\ (p_n)_*\subseteq B_n(M)$. Next, we need a single map
$\phi_s:H\rightarrow Aut(G)$, determined by choosing a
{\it transversal} function $s:H\rightarrow B_n(M)$ [that is, $(p_n)_*\circ
s=id_H$] and
defining $\phi_s(h)$, $h\in H$, to be the automorphism of the normal subgroup
$G$ of $B_n^E(M)$ given by $\phi_s(h)(g)=s(h)^{-1}gs(h)$, $g\in
G$. Note that
neither $\phi _s$ nor $s$ need be homomorphisms, and different choices of $s$
may lead to distinct maps $\phi_s$. $\phi_s$ is called the {\it action} of
$H$ on $G$ associated with $s$. It is true, however, that for any two
transversal functions $s_1$ and $s_2$ the automorphisms $\phi_{s_1}(h)$ and
$\phi_{s_2}(h)$, $h\in H$,
differ only by an inner automorphism of $G$. Thus, there is a unique
map $\phi:H\rightarrow Out(G)$ associated with the extension \ex , where
$Out(G)=Aut(G)/Inn(G)$ is the {\it outer automorphism group} of $G$. Moreover,
$\phi$ is a homomorphism and is called the {\it coupling}\foot{If $G$ is an
abelian group, then $Out(G)=Aut(G)$ and for any choice of $s$, $\phi_s$ is
a homomorphism equal to $\phi$.} of \ex .

If there exists a choice of $s$ which is a homomorphism, then the extension
\ex\  is said to be {\it split} and the above information completely determines
$B_n^E(M)$. In this case, $\phi_s$ is a homomorphism and $B_n^E(M)$ is
isomorphic to the {\it semidirect product} of $G$ by $H$ corresponding to this
action $\phi_s$. If no such choice for $s$ exists, then we also need
information concerning the relevant obstructions. More precisely, suppose that
a certain relation holds between various elements of $H$. Without loss of
generality we can take this relation to be of the form $w=e$, where $w$ is a
product of elements in $H$. Now for a given $s$, the relation $s(w)=e$ need not
hold in $B_n^E(M)$; all we know is that $s(w)=g$, for some $g\in G$. (The
element $g$ depends on both $s$ and $w$.) So suppose $R$ is a set of {\it
defining relations} for $H$, that is, a set of relations from which all others
follow. Then in order to determine the extension $B_n^E(M)$ we need, along with
$G$, $H$, and a fixed $\phi_s$, the set $R^{(s)}$ of relations in $B_n^E(M)$
obtained by ``lifting'' those in $R$ via $s$. Again, if the extension is split,
then $s$ can be chosen such that $s(w)=e$ if $w=e$, and therefore the set
$R^{(s)}$ contains no new information.

Often, the knowledge we possess about $G$ and $H$ is in terms of a
{\it presentation}. We write $G=\ <X_G\ |\ R_G>$ and $H=\ <X_H\ |\ R_H>$ to
denote that $G$ (respectively, $H$) is generated by the set $X_G$
(respectively, $X_H$) subject to the defining relations $R_G$ (respectively,
$R_H$). Given a transversal function $s$, the
above discussion then allows us to construct a presentation of $B_n^E(M)$:
\eqn\pres{B_n^E(M)=\ <X_G,\ s(X_H)\ |\ R_G,\ R_H^{(s)},\ T_s>,}
where
\eqn\rel{T_s=\{ s(y)^{-1}xs(y)=w^{(s)}_{x,y},\ x\in X_G,\ y\in X_H\} .}
The object $w^{(s)}_{x,y}$ is a representation of
$\phi_s(y)(x)\in G$ as a product of elements in $X_G\cup X_G^{-1}$.
The set of relations $T_s$ provides the action of $H$ on $G$
associated with $s$, while $R_H^{(s)}$
gives the required information on the possible obstructions to a splitting.
This formula will prove very useful for us. Finally, we remark that the
obstructions to a splitting homomorphism for \ex\ are related to
certain obstructions to constructing a {\it section} for the map
$p_n$, that is, a continuous map $\omega :Q_n(M)\rightarrow Q_n^E(M)$
with $p_n\circ\omega =id_{Q_n(M)}$. One can try to construct
$\omega$ in stages --- first by finding a section $\omega^{(1)}$ over the
1-skeleton of the $nd$-dimensional manifold $Q_n(M)$, then a section
$\omega^{(2)}$ over the 2-skeleton,
etcetera. At each step there may be an obstruction. If $\omega^{(1)}$ can be
found, then the map $\beta_n:B_n(M)\rightarrow\pi_0(Y)^n$ is trivial
($Im\ (p_n)_*=B_n(M)$). If $\omega^{(2)}$ exists,
there is a splitting homomorphism $s=\omega_*^{(2)}$ for
\ex .\foot{Even if $\omega^{(2)}$ does not exist, there may still be a
splitting $s$ for \ex . However $s$ will not be natural in the sense that it
will have no {\it topological} origin.}
If $\omega^{(3)}$ exists, then the homomorphism
$\alpha_n:\pi_2(Q_n(M))\rightarrow \pi_1(Y)^n$ is also trivial.
Moreover, it is straightforward to show that for $n\geq 2$ and fixed $M$,
a given partial section
$\omega^{(m)}$ for $p_n$ exists if and only if the corresponding
partial section exists for $n=1$ (that is, for $p$).

As an alternative to \ex , we may use the long exact homotopy sequence of the
covering projection $\chi_n:(E^n-\Delta_p)\rightarrow Q_n^E(M)$ to obtain
another extension:
\eqn\ano{\{ e\}\rightarrow \pi_1(E^n-\Delta_p)\buildrel (\chi_n)_*\over
\rightarrow B_n^E(M)\buildrel\gamma_n\over\rightarrow S_n\rightarrow\{ e\} .}
For $Y$ a point, this sequence has often been used to study $B_n(M)$.
There are two simplifications in \ano\ when $d$ = dim $M\geq 3$.
First, one can show in
general that the codimension of $\Delta_p$ in $E^n$ is $d$. So when $d\geq 3$,
this yields $\pi_1(E^n-\Delta_p)=\pi_1(E)^n$.
Moreover, using techniques similar to those developed for the structureless
case \ti , one can show that the extension
\ano\ splits if $d\geq 3$. Thus, in this case, $B_n^E(M)$ can be viewed as a
semidirect product of $\pi_1(E)^n$ by $S_n$; the group $S_n$ acts
by permuting the $n$ factors of $\pi_1(E)$. This semidirect product is
known as the {\it wreath product} and denoted by $\pi_1(E)\wr S_n$.
To go further, one must look at the
long exact sequence of $Y\buildrel i\over\hookrightarrow E\buildrel
p\over\rightarrow M$ to acquire information
about $\pi_1(E)$. It is simply a matter of taste whether one prefers to use
the extension \ex\ or \ano\ to calculate $B_n^E(M)$.
Identical inputs are needed to take advantage of either one.

\newsec{$B_n^F(M)$ for Spin Manifolds ($d\geq 3$)}

We now return to the situation where $M$ is an orientable manifold of
dimension $d\geq 3$, and $E=F(M)$ is the frame bundle of $M$. The fiber
$Y=SO(d)$ of $F(M)$ is path-connected, so $H=Im\ (p_n)_*=B_n(M)$. Moreover,
$\pi_1(SO(d\geq 3))=\IZ_2$ and hence \ex\ becomes
\eqn\ef{\{ e\}\rightarrow\IZ_2^n/Im\ \alpha_n\buildrel i_*\over\rightarrow
B_n^F(M)\buildrel (p_n)_*\over\rightarrow B_n(M)\rightarrow\{ e\}.}
We have $\alpha_n=(\alpha_1)^n\circ\mu_n$ for all $n$ (see Section 2), where
$\alpha_1:\pi_2(M)\to\IZ_2$ can be described as follows. First, an element of
$\pi_2(M)$ may be thought of (up to homotopy) as a sequence of loops $\ell_t$
in $M$ (based at some point $m$), $\ 0\leq t\leq 1$, with
$\ell_0=\ell_1=$ the constant loop. Now consider transporting a frame $v$ at
$m$ along a given $\ell_t$. After completing the loop, it will have been
rotated by some element $R_t\in SO(d)$. Since $R_0=R_1=$ the identity, the
sequence $R_t$ defines a loop in $SO(d)$. The map from $\pi_2(M)$ to
$\pi_1(SO(d))$ obtained in this way is the homomorphism $\alpha_1$.

As noted earlier, much is known about the groups $B_n(M)$ appearing in \ef .
Indeed, using the discussion following \ano\ with $Y$ a point, we can write
$B_n(M)=\pi_1(M)\wr S_n$. (This is not true for $d=2$.)
If $\pi_1(M)=\ <X\ |\ R>$, then $B_n(M)$ is generated by $n$ copies of $X$
along with certain elements $\sigma_i,\ 1\leq i\leq n-1$. These generators
may be considered to be (homotopy classes of) loops in the structureless
particle configuration space $Q_n(M)$. Assume the particles are initially at
the positions $m_1, \dots,m_n$, which can be taken to sit in a $d$-disk
$D\subset M$. Then an element $x^{(i)}$ from the $i$th copy $X^{(i)}$ of $X$
represents a loop in $Q_n(M)$ which takes the particle at
$m_i$ around a loop in $M$ (avoiding all other particles) in the homotopy
class $x\in X\subseteq\pi_1(M)$. The element $\sigma_i$ represents the local
interchange in $D$ of the particle at $m_i$ with that at $m_{i+1}$. The
defining relations for $B_n(M)$ will include $n$ copies of $R$, one for
each $X^{(i)}$, as well as
\eqn\rela{\eqalign{x^{(i)}y^{(j)}&=y^{(j)}x^{(i)}\cr
x^{(i+1)}\sigma_i&=\sigma_i x^{(i)}\cr
x^{(j)}\sigma_i&=\sigma_i x^{(j)}\cr
\sigma_i^2&=e\cr
\sigma_i\sigma_{i+1}\sigma_i&=\sigma_{i+1}\sigma_i\sigma_{i+1}\cr
\sigma_i\sigma_j&=\sigma_j\sigma_i\cr}
\qquad\eqalign{&1\leq i,j\leq n;\ i\neq j,
\cr &1\leq i\leq n-1,\cr
&1\leq i\leq n-1;\ 1\leq j\leq n;\ j\neq i,i+1,\cr
&1\leq i\leq n-1,\cr
&1\leq i\leq n-2,\cr
&1\leq i,j\leq n-1;\ |i-j|\geq 2,\cr}}
where $x^{(i)},y^{(i)}\in X^{(i)}$. A given $X^{(i)}$ generates a subgroup
isomorphic to $\pi_1(M)$, and the $\sigma_i$'s generate an $S_n$ subgroup.
Clearly we have $B_1(M)=\pi_1(M)$.

To compute the extension in \ef , we still need to know the connecting
homomorphism $\alpha_n$ as well as how the relations in $B_n(M)$
above lift to $B_n^F(M)$. As mentioned previously, there exists a splitting
homomorphism for \ef\  if there is a partial section $\omega^{(2)}$ for
${p:F(M)\rightarrow M}$. If $\omega^{(2)}$ can be extended to $\omega^{(3)}$,
then $\alpha_n$ is trivial. The obstructions to finding a section for $p$
are related to various characteristic classes $t_i\in
H^i(M;\pi_{i-1}(SO(d))),\ i\geq 1$. The map $\omega^{(m)}$ exists for $p$
if and only if $t_i=0$ for all $1\leq i\leq m$. We state a few basic
facts about these classes. Since $SO(d)$ is path-connected, $t_1=0$ for any
$M$. Next, the element $t_2\in H^2(M;\IZ_2)$ is the {\it second Stiefel-Whitney
class} of the tangent bundle $\tau_M$ of $M$, usually denoted by $w_2$.
The class of orientable spaces which have $w_2=0$ are called spin
manifolds. This is because they are precisely the manifolds for which the
$SO(d)$-bundle $F(M)$ can be extended to a principal $Spin(d)$-bundle over $M$,
where $Spin(d)$ is the double cover of $SO(d)$. Hence, spinors can be
unambiguously defined on these spaces. Finally, $t_3=0$ since $\pi_2(SO(d))$ is
trivial. Thus, we can construct the partial section $\omega^{(3)}$ for $p$ if
and only if $M$ is a spin manifold. So for
these spaces $M_{spin}$ we have a split extension
\eqn\spin{\{ e\}\rightarrow\IZ_2^n\buildrel i_*\over\rightarrow
B_n^F(M_{spin})\buildrel (p_n)_*\over\rightleftharpoons
B_n(M_{spin})\rightarrow\{ e\}.}
All that remains here is to determine the coupling of \spin . To this end,
let us choose as our basepoint in $Q_n^F(M)$ the configuration where the $n$
identical spinning particles are located at the
points $m_1,\dots ,m_n\in D\subset M_{spin}$. We must also pick a specific
frame $v_i$ above each $m_i$. If we assume
$\pi_1(M_{spin})=\ <X\ |\ R>$, then $B_n^F(M_{spin})$ in \spin\ is generated
by $X^{(i)}, 1\leq i\leq n$,
and $\sigma_j,\ 1\leq j\leq n-1$, as for $B_n(M_{spin})$, along with
elements $r_k,\ 1\leq k\leq n$. Here $r_k$ represents a 2$\pi$-rotation of
the frame $v_k$ above $m_k$; the $r_k$'s generate the subgroup
$G=i_*(\IZ_2^n)$.
The defining relations for $B_n^F(M_{spin})$ consist of $n$
copies of $R$, the relations in \rela , as well as
\eqn\coup{\eqalign{r_i^2&=e\cr r_ir_j&=r_jr_i\cr x^{(i)}r_j&=r_jx^{(i)}\cr
r_{i+1}\sigma_i&=\sigma_ir_i\cr
r_j\sigma_i&=\sigma_ir_j\cr}\qquad\eqalign{&1\leq i\leq n,\cr
&1\leq i,j\leq n,\cr &1\leq
i,j\leq n,\cr &1\leq i\leq n-1,\cr &1\leq i\leq n-1;\ 1\leq j\leq n;\ j\neq
i,i+1,\cr}}
where $x^{(i)}\in X^{(i)}$. The first two sets of equations in \coup\ give
relations which hold in $G$. It is not hard to convince oneself that the
remaining three sets hold in $B_n^F(M_{spin})$, and that they provide the
action
of $H=B_n(M_{spin})$ on $G$. It is worth mentioning that the $\sigma_i$'s,
along with the set $X^{(1)}$ and the element $r_1$, are all that's needed
to generate $B_n^F(M_{spin})$. However, eliminating the other generators
from the above presentation leads to a more cumbersome set of relations.
As a special case we have
$B_1^F(M_{spin})=\pi_1(F(M_{spin}))=\IZ_2\times\pi_1(M_{spin})$,
and hence for any $n$
we can write $B_n^F(M_{spin})={(\IZ_2\times\pi_1(M_{spin}))\wr S_n}$.
Our results for $B_n^F(M_{spin})$ apply,
in particular, to closed, orientable 3-manifolds $M^{(3)}$
since they are all spin manifolds.\foot{Moreover, these spaces are
{\it parallelizable} (that is, $F(M^{(3)})=SO(3)\times M^{(3)}$)
because here $\omega^{(3)}$ is a full section for $p$, and sectioned principal
bundles are homeomorphic to a product. Every closed, orientable 2-manifold
is a spin manifold as well. However their treatment is somewhat
different as we shall see in Section 5.}
It is interesting to note that if $M_1$ and $M_2$ are two spin manifolds
(of dimension three or more)
with the same fundamental group, then $B_n^F(M_1)=B_n^F(M_2)$.

Some common examples of spin manifolds are the Euclidean spaces $\IR^d$ and the
spheres $S^d$, as well as the real projective spaces $\IR\IP^{4m+3}$
and the complex projective spaces $\IC\IP^{2m+1}$. By the final
remark of the preceding paragraph we have $B_n^F(\IR^d)=B_n^F(S^d)=
B_n^F(\IC\IP^{2m+1})$, for $d\geq 3$ and $m\geq 1$, since these spaces are all
simply connected and have dimension at least 3 (recall that dim $\IC\IP^m=2m$).
For these manifolds the set $X$ is empty and $B_n^F$ is just the wreath product
$\IZ_2\wr S_n$ generated by the $\sigma_i$'s and the $r_j$'s. This group has
$n\cdot 2^n$ elements. It is easy to check that the one-dimensional unitary
representations of $\IZ_2\wr S_n$, $n\geq 2$, are given
by the four possible combinations $\sigma_i=\pm 1,\ r_j=\pm 1$ ($1\leq i\leq
n-1,\ 1\leq j\leq n$). (If $n=1$, then the group is a single $\IZ_2$
generated by $r_1$ and there are only two IUR's, namely, $r_1=\pm 1$.)
In the corresponding scalar quantum theories, the
particles obey Bose (respectively, Fermi) statistics if $\sigma_i=1$
(respectively, $-1$), and they have integral (respectively, half-integral)
spin if $r_j=1$ (respectively, $-1$).
Note that there exist theories which violate the usual spin-statistics
connection, since we can choose the $\sigma$'s and $r$'s to have opposite
signs.
Since $\IZ_2\wr S_n$ is nonabelian for $n\geq 2$, there will also be higher
dimensional IUR's in these cases and hence nonscalar quantizations. As an
example, consider the case $n=2$. The group $\IZ_2\wr\IZ_2$ is isomorphic to
the dihedral group of order 8. Along with the four IUR's of dimension one
described above, it has a two-dimensional IUR determined by
\eqn\dih{\sigma_1=\pmatrix{1&0\cr 0&-1\cr},\qquad r_2=\pmatrix{0&1\cr 1&0\cr}.}
(The 2$\pi$-rotation $r_1$ is given by $\sigma_1^{-1}r_2\sigma_1$.) In the
associated quantum theory, the two particles obey a type of
``half Bose - half Fermi'' statistics which has been called ambistatistics.
In the past, ambistatistics has been obtained for structureless particles
either on nonsimply connected spaces \ti\ug\jmr , or on simply connected spaces
in the presence of spectators which effectively create noncontractible loops
\ref\pcb{L.~Brekke, A.~Falk, S.~J.~Hughes and T.~D.~Imbo, Phys. Lett.
{\bf B271} (1991) 73.}\ref\gauge{L.~Brekke, H.~Dykstra, A.~Falk and T.~D.~Imbo,
Phys. Lett. {\bf B304} (1993) 127.}. Similar quantizations have also been
found for extended objects such as identical {\it geons} in quantum gravity
\ref\geon{C.~Aneziris, A.~P.~Balachandran, M.~Bordeau, S.~Jo, T.~R.~Ramadas and
R.~D.~Sorkin, Int. J. Mod. Phys. {\bf A4} (1989) 5459.}, topological solitons
in certain nonlinear sigma models \ref\lbti{L.~Brekke and T.~D.~Imbo, Int. J.
Mod. Phys. {\bf A7} (1992) 2589.}\ref\ltip{L.~Brekke and T.~D.~Imbo, in
preparation.}, and strings in mechanics \ref\string{C.~Aneziris,
A.~P.~Balachandran, L.~Kauffman and A.~M.~Srivastava, Int. J. Mod. Phys.
{\bf A6} (1991) 2519\semi C.~Aneziris, Mod. Phys. Lett. {\bf A7} (1992)
3789\semi A.~Brownstein and T.~D.~Imbo, in preparation.} and certain gauge
theories \gauge . In \dih , however, we have obtained ambistatistics for
pointlike particles on a simply connected manifold with no spectators by
introducing an internal spin degree of freedom. We also see that these
particles possess what may be called {\it ambispin}.

For three or more particles we can construct IUR's leading to parastatistics.
These representations, when restricted to the $\sigma_i$'s, yield
IUR's of $S_n$ of dimension two or more.
For instance, $\IZ_2\wr S_3$ has two IUR's of dimension two given by
\eqn\thr{\sigma_1=\pmatrix{1&0\cr 0&-1},\qquad \sigma_2=\pmatrix{-1/2&
\sqrt{3}/2\cr\sqrt{3}/2&1/2},\qquad r_3=\pm\pmatrix{1&0\cr 0&1}.}
The matrices for $\sigma_1$ and $\sigma_2$ generate the two-dimensional IUR
of $S_3$. So both of the representations in \thr\ give parastatistics for
the three particles. If we choose the plus sign for $r_3$ ($=r_2=r_1$), then
these paraparticles have integral spin. The minus sign yields half-integral
spin. There are also four IUR's of $\IZ_2\wr S_3$ of dimension three. They
give rise to other types of exotic statistics. Similar results hold for
$n\geq 4$.

For nonsimply connected spin manifolds $M_{spin}$ ($d\geq 3$), the groups
$B_n^F(M_{spin})$ are ``larger'' than above, and their IUR's
correspondingly more complex.
However, $\IZ_2\wr S_n$ is always a homomorphic image of $B_n^F(M_{spin})$
(just set the generators in each $X^{(i)}$ equal to $e$). As a consequence,
every IUR of $\IZ_2\wr S_n$ can be naturally viewed as an IUR of $B_n^F
(M_{spin})$. If we let $\xi$ denote the homomorphism from $B_n^F(M_{spin})$
onto $\IZ_2\wr S_n$, then the IUR of $B_n^F(M_{spin})$ associated with the IUR
$\rho$ of $\IZ_2\wr S_n$ is simply $\rho\circ\xi$. Hence, the representations
described in the simply connected case above, as well as their quantum
mechanical interpretations, are still relevant for a general $M_{spin}$. There
may, of course, be further IUR's. We conclude this section by noting one
implication of the above discussion which will be of interest to us later.
Namely, consider $n$ identical particles moving on an arbitrary spin manifold
of dimension three or more. Then, for any $n$, there exist scalar quantizations
of this system where the particles have half-integral spin.

\newsec{$B_n^F(M)$ for Orientable Nonspin Manifolds ($d\geq 3$)}

We now turn our attention to the situation where $M$ is an orientable, nonspin
manifold with $d\geq 3$. The exact sequence \ef\ is still valid here.
One major difference from the spin manifold case is that the homomorphism
$\alpha_n$ is no longer trivial in general. We will see
that if $\alpha_n$ is not trivial, then it is onto and thus
$B_n^F(M)=B_n(M)$. But if it {\it is}
trivial, another difference is that the extension \ef\ will not be split.
Therefore we must understand exactly when $\alpha_n$ is trivial, and if so how
the relations in $B_n(M)$ lift to $B_n^F(M)$. The performance of this task
requires studying in more detail the obstructions to defining a spin structure
on $M$.

As mentioned earlier, an orientable manifold $M$ is spin if and only if the
second Stiefel-Whitney class $w_2\in H^2(M;\IZ_2)$ is trivial. However,
there are two useful alternatives to this cohomological characterization of the
obstruction to a spin structure. One involves the first two homotopy groups of
$M$, while the other uses the first two homology groups. To describe the
homotopy-theoretic alternative, we exhibit $\pi_1(F(M))$ as an
extension by using \ef\ with $n=1$:
\eqn\eft{\{ e\}\rightarrow\IZ_2/Im\ \alpha_1\buildrel i_*\over\rightarrow
\pi_1(F(M))\buildrel p_*\over\rightarrow \pi_1(M)\rightarrow\{ e\}.}
$M$ is said to have a $\pi_1$-{\it obstruction} to a spin structure if \eft\
does not split. (Note that this implies that $\alpha_1$ is trivial.) We say
that $M$ has a $\pi_2$-{\it obstruction} to a spin structure if $\alpha_1$ is
nontrivial.

Every orientable, nonspin manifold has one or the other (but by definition not
both) of these obstructions. Some well known examples are the complex
projective spaces $\IC\IP^{2m}$ and the real projective spaces $\IR\IP^{4m+1}$,
$m\geq 1$. The former spaces are simply connected, but possess a
$\pi_2$-obstruction to a spin structure ($\pi_2(\IC\IP^{2m})=\IZ$). The latter
spaces have trivial $\pi_2$, but possess a $\pi_1$-obstruction (their
fundamental group is $\IZ_2$). Note that the product of a space with a
$\pi_1$-obstruction and a space having a $\pi_2$-obstruction, like
${\IR\IP^5\times\IC\IP^2}$, has a $\pi_2$-obstruction. In general, information
about these two obstructions allows us to determine $B_n^F(M)$. For example,
since (by definition) $\alpha_1$ is an epimorphism for spaces with a
$\pi_2$-obstruction to a spin structure, we see that each $\alpha_n$,
$n\geq 2$, is also onto for these manifolds. (Recall that $\alpha_n=
(\alpha_1)^n\circ\mu_ n$, and that $\mu_n$ is onto for $d\geq 3$.) Thus,
we have:
\vskip 4pt
\noindent
{\it If an orientable manifold $M$ has a $\pi_2$-obstruction to a spin
structure, then $B_n^F(M)=B_n(M)$ for all $n\geq 1$.}
\vskip 4pt
\noindent
As a consequence, in the quantum theories associated with IUR's of
$B_n^F(M)$ the $n$ identical particles must have integral spin (since the
2$\pi$-rotations are homotopically trivial in $Q_n^F(M)$). There will
still be, in general, a wide spectrum of statistics for the particles in these
quantum theories. For instance, we are assured of at least one IUR yielding
Fermi statistics. As an example, consider the spaces $\IC\IP^{2m},\ m\geq 1$,
described above. It is easy to show that $B_n^F(\IC\IP^{2m})=S_n$ for any
$m$. We thus obtain Bose, Fermi and parastatistical quantizations, but the
identical particles in each of these theories have
integral spin. In Section 8 we will see how half-integral spin may be
obtained on $\IC\IP^{2m}$ even though it does not allow ordinary spinors.

What about the case when $M$ has a $\pi_1$-obstruction?
Here the map $\alpha_n$ is trivial and we have
\eqn\pon{\{ e\}\rightarrow\IZ_2^n\buildrel i_*\over\rightarrow B_n^F(M)
\buildrel (p_n)_*\over\rightarrow B_n(M)\rightarrow\{ e\}.}
In this case, $B_n^F(M)$ has the ``same'' set of generators as for $M$ a spin
manifold. Further, \pon\ has the same coupling as \spin\ so that the relations
in \coup\ hold here as well. The extension \pon , however, {\it does not
split}. So there are relations in $B_n(M)$ which get modified when
lifted to $B_n^F(M)$. From the discussion in the last paragraph of Section 2,
it can be deduced that the relations in \rela\ go over without change
to $B_n^F(M)$. It is only the $n$ copies $R^{(i)}$ of the set $R$ of relations
given in the presentation $\pi_1(M)=\ <X\ |\ R>$ that must be modified. For
each $1\leq i\leq n$, the new set ${\tilde R}^{(i)}$ of lifted relations can
be obtained as follows. First, choose a representative loop $\ell_x$ for each
homotopy class $x\in X$. (We assume all loops are based at $m_i\in M$.) This
allows us to find a representative for each {\it word} in these generators,
that is, for each string of products of elements in $X\cup X^{-1}$. In
particular, we can construct a representative $\ell_w$ for each word $w$ such
that $w=e$ is a relation in $R$. We may then lift $\ell_w$ to a loop ${\tilde
\ell}_w$ in $F(M)$ and deform it so that it lies completely in the $SO(d)$
fiber over $m_i$. If this procedure yields a loop which is homotopically
trivial in $SO(d)$, then the relation $w=e$ carries over to the set
${\tilde R}^{(i)}$. On the other hand, if this deformation of ${\tilde \ell}_w$
lies in the class of the 2$\pi$-rotation, then the new relation $w=r_i$
replaces $w=e$. This gives us ${\tilde R}^{(i)}$. To recap, let $M$ be an
orientable manifold ($d\geq 3$) with a $\pi_1$-obstruction to a spin structure.
Also let $\pi_1(M)=\ <X\ |\ R>$. Then $B_n^F(M)$ is generated by the $n$ copies
$X^{(i)}$ of the set $X$, along with the $n-1$ exchanges $\sigma_i$ and the $n$
rotations $r_j$. The defining relations consist of the sets ${\tilde R}^{(i)}$
just described, as well as the relations in \rela\ and \coup .

As an example, consider the spaces $\IR\IP^{4m+1},\ m\geq 1$. For any $m$
we have $\pi_1(\IR\IP^{4m+1})=\ <x\ |\ x^2=e>\ =\IZ_2$, and for each $1\leq
i\leq n$ the set ${\tilde R}^{(i)}$ contains the single relation
$x^2=r_i$. So the group $B_1^F(\IR\IP^{4m+1})=\pi_1(F(\IR\IP^{4m+1}))$ is
isomorphic to $\IZ_4$, where the 2$\pi$-rotation  $r$ is the square of the
generator $x$. There are four IUR's of $\IZ_4$ given by $x=\pm 1$ and
$x=\pm i$. In the quantizations of the one-particle system corresponding to
the first two IUR's, the particle has integral spin ($x^2=r=1$). In the
remaining two, the particle has half-integral spin ($x^2=r=-1$). Of course
for higher $n$ we have $B_n^F(\IR\IP^{4m+1})=\IZ_4\wr S_n$, and similar
statements hold about the allowed spins for the $n$ identical particles.
More generally, one can prove that there exists an IUR of $B_n^F(M)$ such
that the corresponding quantum theory yields half-integral spin if and only
if $M$ has no $\pi_2$-obstruction. It is curious that half-integral spin is
allowed on spaces with a $\pi_1$-obstruction even though they do not admit
ordinary spinors. In many cases, this is due to the existence of a generalized
spin structure on the manifold called a ${\rm spin}_c$ structure. This will be
discussed in more detail in Section 8.

An alternative way of describing the possible obstructions to a spin structure
for an orientable manifold $M$ utilizes the first two homology groups of $M$.
More specifically, consider the following commutative diagram:
\eqn\cdo{\matrix{&\pi_2(M) &\buildrel\alpha_1\over\longrightarrow &\IZ_2
&\buildrel i_*\over\longrightarrow &\pi_1(F(M)) &\buildrel p_*\over
\longrightarrow &\pi_1(M) &\to &\{ e\}\cr
&\Big\downarrow &\ &id\Big\downarrow\phantom{id} &\
&\Big\downarrow &\ &\Big\downarrow &\ &\ \cr
&H_2(M) &\buildrel\tau\over\longrightarrow &\IZ_2 &\buildrel i_*\over
\longrightarrow &H_1(F(M))
&\buildrel p_*\over\longrightarrow &H_1(M) &\to
&\{ e\} .\cr}}
The bottom row is the {\it Serre exact homology sequence} of the frame bundle
of $M$, and the homomorphism $\tau$ is known as the {\it transgression}.
Each of the vertical maps is the {\it Hurewicz homomorphism}. In particular,
the second one from the left, namely $id$, is the identity map. We say that $M$
has an $H_1$-{\it obstruction} to a spin structure if the extension
\eqn\efth{\{ e\}\rightarrow\IZ_2/Im\ \tau\buildrel i_*\over\rightarrow
H_1(F(M))\buildrel p_*\over\rightarrow H_1(M)\rightarrow\{ e\}}
does not split. $M$ is said to possess an $H_2$-{\it obstruction} if $\tau$ is
nontrivial.

Again, every orientable, nonspin manifold has one or the other (but not both)
of these obstructions. From the commutativity of \cdo\ we see that spaces with
a $\pi_2$-obstruction necessarily have an $H_2$-obstruction. Thus, for example,
each of the spaces $\IC\IP^{2m}$ above has an $H_2$-obstruction to a spin
structure. The spaces $\IR\IP^{4m+1}$ discussed earlier have an
$H_1$-obstruction since $H_2$ of each of these spaces is trivial. However,
there is no general connection between $\pi_1$- and $H_1$-obstructions. Indeed,
there exist orientable manifolds which have a $\pi_1$- and an
$H_2$-obstruction to a spin structure. The simplest example we know of is the
5-dimensional closed, flat Riemmanian manifold constructed in
\ref\frm{L.~Auslander and R.~H.~Szczarba, Ann. Math. {\bf 76} (1962) 1.}. This
nonspin manifold, which we denote by $M_5$, has a nonabelian, torsion-free
fundamental group and is {\it aspherical}\foot{Indeed all closed, flat
Riemmanian manifolds are aspherical, and all finite-dimensional aspherical
spaces must have torsion-free fundamental groups \ref\char{L.~S.~Charlap,
{\it Bieberbach Groups and Flat Manifolds} (Springer-Verlag, New York,
1986).}.} (that is, $\pi_n(M_5)$ is trivial for $n\geq 2$). Thus, it must have
a $\pi_1$-obstruction to a spin structure. However, $M_5$ can be shown to
possess an $H_2$-obstruction as well \ref\dui{M.~J.~Dugan and T.~D.~Imbo, in
preparation.}. Generalizing the methods of \frm , the author of
\ref\vas{A.~T.~Vasquez, J. Diff. Geom. {\bf 4} (1970) 367.} constructed closed,
flat Riemmanian manifolds of any odd dimension $\geq 5$ which have $w_2\neq 0$.
All of these spaces can be shown to have a $\pi_1$- and an $H_2$-obstruction
to a spin structure. Note that if a manifold $M$ has a $\pi_1$- and an
$H_2$-obstruction, then the $2\pi$-rotation $r\in\pi_1(F(M))$ is nontrivial
and lies in the {\it commutator subgroup}. Thus, half-integral spin
quantizations for a particle on $M$ exist, but they are necessarily nonscalar.
(Similar results hold for $n$ particles on $M$.) In general, there exists
a {\it one-dimensional} IUR of $B_n^F(M)$ yielding half-integral spin for $n$
identical particles on a manifold $M$ if and only if $M$ has no
$H_2$-obstruction to a spin structure.

We should point out that the notion of what we call here a $\pi_2$-obstruction
to a spin structure has been discussed in the past (see, for example,
\ref\war{S.~W.~Hawking and C.~N.~Pope, Phys. Lett. {\bf B73} (1978) 42\semi
L.~Castellani, L.~J.~Romans and N.~P.~Warner, Nucl. Phys. {\bf B241} (1984)
429.}\ref\pen{R.~Penrose and W.~Rindler, {\it Spinors and Space-Time}, Vol. 2
(Cambridge University Press, Cambridge, 1986).}). The authors of \pen\ also
discuss an obstruction to defining spinors having its origin in the fundamental
group of the manifold $M$ in question. This is similar to our
$\pi_1$-obstruction. They further claim that if $\pi_1(M)$ has no elements of
even order, then there is no such obstruction. However the examples of \frm\
and \vas\ clearly show that noncontractible loops in $M$ can be the source of
an obstruction to a spin structure even if $\pi_1(M)$ is torsion-free. Finally,
as far as we can tell, the $H_1$- and $H_2$-obstructions defined above have
not been discussed previously.

\newsec{$B_n^F(M)$ for Orientable Surfaces}

In the previous two sections we have calculated the groups $B_n^F(M),\
n\geq 1$, for an arbitrary orientable manifold of dimension three or more.
The computations for an orientable 2-manifold $M^{(2)}$ are more difficult.
The additional complications have three main sources. First, for any bundle $E$
over $M^{(2)}$, the codimension of $\Delta_p$ in $E^n$ is only two and hence
$\pi_1(E^n-\Delta_p)\neq\pi_1(E)^n$ in general. The low codimension of
$\Delta_p$ further implies that the homomorphism
$\mu_n:\pi_2(Q_n(M^{(2)}))\rightarrow\pi_2(M^{(2)})^n$ is not necessarily onto.
Second, unlike in higher dimensions, the extension \ano\ generally does not
split. In particular, the exchanges $\sigma_i$ may no longer square to the
identity. This is the reason why fractional statistics can occur in two spatial
dimensions. Finally, we have that $\pi_1(SO(2))=
\IZ$ and not $\IZ_2$. This is why fractional spin is allowed in two dimensions
(the 2$\pi$-rotations $r_i$ need not square to the identity). The following
exact sequence replaces \ef :
\eqn\eft{\{ e\}\rightarrow\IZ^n/Im\ \alpha_n\buildrel i_*\over\rightarrow
B_n^F(M^{(2)})\buildrel (p_n)_*\over\rightarrow B_n(M^{(2)})\rightarrow
\{ e\}.}
If the characteristic class $t_2\in H^2(M^{(2)};\IZ)$ of the tangent bundle
$\tau_{M^{(2)}}$ vanishes, then $p:F(M^{(2)})\rightarrow M^{(2)}$ has a full
section, implying that $\alpha_n$ is trivial and the extension \eft\ splits.
(Note that $t_1$ is still zero.) But $t_2$ lives in the second {\it integral}
cohomology group and so is no longer the second Stiefel-Whitney class;
it is the {\it Euler class}. Hence, $t_2$ need not vanish
even if $M^{(2)}$ is a spin manifold. It is worth mentioning here that if
$\pi_1(M^{(2)})=\ <X\ |\ R>$, then, as for $d\geq 3$, $B_n^F(M^{(2)})$ can be
generated by $n$ copies of $X$ along with the $\sigma_i$'s and $r_i$'s.
The set of defining relations, however, is much more cumbersome.\foot{When
working in two spatial dimensions, it is important to make our orientation
conventions for the $\sigma$'s and $r$'s explicit since they may not square
to the identity. We always choose the $\sigma$'s to be ``clockwise'' exchanges,
and the $r$'s similarly to be clockwise rotations.}

Things simplify for flat surfaces $M_{flat}^{(2)}$,
since they are parallelizable. Examples are
the plane $\IR^2$, the torus $T^2$, and either of these with an
arbitrary number of punctures (like the cylinder which is homeomorphic to
$\IR^2-\{ 0\}$). For these spaces $t_2=0$, so that $B_n^F(M_{flat}^{(2)})$
is a semidirect product of $\IZ^n$ by $B_n(M_{flat}^{(2)})$. The last four
sets of equations in \coup\ hold in $B_n^F(M_{flat}^{(2)})$, with the last
three giving the coupling of the above semidirect product. To these relations,
one need only add those for $B_n(M_{flat}^{(2)})$ to obtain a complete
presentation of $B_n^F$. For example, the last two sets of equations in \rela\
are a full set of defining relations for the classical braid group\foot{These
local relations clearly hold in the braid group of any other 2-manifold, but
there are, most often, numerous others as well.} $B_n(\IR^2)$, and using this
we obtain $B_n^F(\IR^2)$ \ref\ko{K.~H.~Ko and L.~Smolinsky, Proc. Amer. Math.
Soc. {\bf 115} (1992) 541.}. Its one-dimensional IUR's are given by $\sigma_i=
e^{{\rm i}\theta}$, $r_j=e^{{\rm i}\phi}$, for $0\leq\theta,\ \phi <2\pi$ and
all $i$ and $j$. Thus we obtain the full spectrum of fractional spin and
statistics for $n$ identical particles on $\IR^2$. Higher-dimensional IUR's
of $B_n^F(\IR^2)$ give rise to other types of exotic spin and statistics.
The ordinary braid groups $B_n$ of the cylinder \ref\cyl{D.~L.~Goldsmith, Math.
Scand. {\bf 50} (1982) 167\semi Y.~Hatsugai, M.~Kohmoto and Y.-S.~Wu, Phys.
Rev. {\bf B43} (1991) 10761.} and the torus \ref\tor{J.~S.~Birman, Comm. Pure
Appl. Math. {\bf 22} (1969) 41.}\ref\fr{Y.~Ladegaillerie, Bull. Sci. Math.
{\bf 100} (1976) 255.} are also well known, and the above procedure can be used
to determine $B_n^F$ for these spaces.

The groups $B_n$ of any closed, orientable surface are known as well (for
$S^2$ see \ref\fvb{E.~Fadell and J.~Van Buskirk, Duke Math. Jour. {\bf 29}
(1962) 243.}, for all higher genus surfaces see \tor\fr ). But here, except for
the torus, $t_2\neq 0$ and much more work has to be done to calculate $B_n^F$.
For a closed, orientable surface $M_g^{(2)}$ of genus $g$, we have
$H^2(M_g^{(2)};\IZ)=\IZ$ and we can choose $t_2={2(1-g)}$. If $g\geq 1$ then
$\pi_2(M_{g\geq 1}^{(2)})=\{ e\}$, which has been shown to imply that $\pi_2
(Q_n^F(M_{g\geq 1}^{(2)}))=\{ e\}$ for all $n\geq 1$ \bgt . This yields
\eqn\yie{\{ e\}\rightarrow\IZ^n\buildrel i_*\over\rightarrow B_n^F(M_{g\geq 1}
^{(2)})\buildrel (p_n)_*\over\rightarrow B_n(M_{g\geq 1}^{(2)})\rightarrow
\{ e\}.}
The group $B_n(M_{g\geq 1}^{(2)})$ is generated by the exchanges $\sigma_i,\
1\leq i\leq n-1$, along with elements $\rho_l,\ \tau_l,\ 1\leq l\leq g$. The
$\rho$'s (respectively, $\tau$'s) correspond to taking the particle at $m_1$
around the $g$ meridianal (respectively, longitudinal) homology cycles in
$M_{g\geq 1}^{(2)}$ in the manner described in \fr . The defining relations
among these generators are given by the last two sets of equations in \rela\
along with
\eqn\mod{\sigma_1\dots\sigma_{n-1}^2\dots\sigma_1\tau_g\tau_{g-1}\dots\tau_1
(\rho_1^{-1}\tau_1^{-1}\rho_1)\dots(\rho_g^{-1}\tau_g^{-1}\rho_g)=e,}
and
\eqn\oth{\eqalign{\sigma_i\rho_l&=\rho_l\sigma_i\cr \sigma_i\tau_l&=\tau_l
\sigma_i\cr \sigma_1\rho_m\sigma_1\rho_l&=\rho_l\sigma_1\rho_m\sigma_1\cr
\sigma_1\tau_m\sigma_1^{-1}\tau_l&=\tau_l\sigma_1\tau_m\sigma_1^{-1}\cr
\sigma_1\tau_l\sigma_1\tau_l&=\tau_l\sigma_1\tau_l\sigma_1\cr
\sigma_1\tau_m\sigma_1^{-1}\rho_l&=\rho_l\sigma_1\tau_m\sigma_1^{-1}\cr
\sigma_1\rho_m\sigma_1\tau_l&=\tau_l\sigma_1\rho_m\sigma_1\cr
\sigma_1^{-1}\rho_l\sigma_1\tau_l&=\tau_l\sigma_1\rho_l\sigma_1\cr}\qquad
\eqalign{&2\leq i\leq n-1;\ 1\leq l\leq g,\cr
&2\leq i\leq n-1;\ 1\leq l\leq g,\cr
&m\geq l;\ 1\leq l,m\leq g,\cr
&m>l;\ 1\leq l,m\leq g,\cr
&1\leq l\leq g,\cr
&m>l;\ 1\leq l,m\leq g,\cr
&m>l;\ 1\leq l,m\leq g,\cr
&1\leq l\leq g.}}

The relation \mod\ gets modified when lifted to $B_n^F(M_{g\geq 1}^{(2)})$.
The new relation reads \tte\apb\ (see also \jmr )
\eqn\new{\sigma_1\dots\sigma_{n-1}^2\dots\sigma_1\tau_g\tau_{g-1}\dots\tau_1
(\rho_1^{-1}\tau_1^{-1}\rho_1)\dots(\rho_g^{-1}\tau_g^{-1}\rho_g)
=r_1^{2(g-1)}.}
By contrast, the relations in \oth\ remain valid when lifted. To complete the
presentation of $B_n^F(M_{g\geq 1}^{(2)})$ we add to \oth\ and \new\ the
following:
\eqn\fol{\eqalign{
\sigma_i\sigma_{i+1}\sigma_i&=\sigma_{i+1}\sigma_i\sigma_{i+1}\cr
\sigma_i\sigma_j&=\sigma_j\sigma_i\cr
r_ir_j&=r_jr_i\cr
r_{i+1}\sigma_i&=\sigma_ir_i\cr
r_j\sigma_i&=\sigma_ir_j\cr
\rho_lr_j&=r_j\rho_l\cr \tau_lr_j&=r_j\tau_l\cr}
\qquad\eqalign{&1\leq i\leq n-2,\cr
&1\leq i,j\leq n-1;\ |i-j|\geq 2,\cr
&1\leq i,j\leq n,\cr
&1\leq i\leq n-1,\cr &1\leq i\leq n-1;\ 1\leq j\leq n;\ j\neq i,i+1,\cr
&1\leq l\leq g;\ 1\leq j\leq n,\cr &1\leq l\leq g;\ 1\leq j\leq n,\cr}}
which are the relevant portions of \rela\ and \coup . For $n=1$, set each
$\sigma_i=e$ in \new . (To obtain $B_n^F$ for
the {\it connected sum} $\IR^2\# M_{g\geq 1}^{(2)}$, or equivalently
$M_{g\geq 1} ^{(2)}$ with a single point removed, simply ignore \new\ in the
presentation of $B_n^F(M_{g\geq 1}^{(2)})$). For a partial treatment of the
representation theory of these groups, and the attendant consequences for the
fractional spin and statistics of identical particles on these spaces, see
\jmr\tte . These systems will also possess quantizations in which the
particles obey fractional ambistatistics and have fractional ambispin.

The case $M^{(2)}=S^2$ (that is, $g=0$) is a bit different. $F(S^2)$ is
homeomorphic to $\IR\IP^3$ and thus $B_1^F(S^2)=\pi_1(F(S^2))=\IZ_2$.
This group is generated by the single 2$\pi$-rotation, which here {\it does}
square to the identity. So fractional spin is not allowed for the particle.
For any $n\geq 2$, the following genus zero version of \new\ holds in
$B_n^F(S^2)$ \tte :
\eqn\ne{\sigma_1\dots\sigma_{n-1}^2\dots\sigma_1=r_1^{-2}.}
As $S^2$ is simply connected, there is no analog of the
relations in \oth . To complete the presentation of $B_n^F(S^2)$, we simply
add the first five sets of relations in \fol . This can be proven using,
among other things, the
results in \ref\fad{E.~Fadell, Duke Math. J. {\bf 29} (1962) 231.}.
In particular, we note that for $n\geq 3$
we have $\pi_2(Q_n(S^2))=\{ e\}$, while $\pi_2(Q_2(S^2))=\IZ$ and the
homomorphism $\alpha_2$ is nontrivial.

For any {\it open} orientable surface $M_{open}^{(2)}$, it is straightforward
to show that $\pi_2(M_{open}^{(2)})=\{ e\}$. Again, this implies that
$\pi_2(Q_n(M_{open}^{(2)}))=\{ e\}$ for all $n\geq 1$.
Hence, the homomorphisms $\alpha_n$ are
trivial for these spaces and there is an exact sequence for $B_n^F(M_{open}
^{(2)})$ analogous to \yie\ for $M_{g\geq 1}^{(2)}$. If $M_{open}^{(2)}$ is
flat, we have already given a characterization of this extension in terms of
$B_n(M_{open}^{(2)})$. All other cases must be handled individually,
since there are no simple classification theorems for open 2-manifolds.

\newsec{$B_n^F(M)$ for Nonorientable Spaces}

If $M_{NO}$ is a {\it nonorientable} manifold of dimension $d\geq 2$, then the
fiber of $F(M_{NO})$ is the full orthogonal group $O(d)$. For any $d$ we have
$\pi_0(O(d))=\IZ_2$, and the map $\beta_n:B_n(M_{NO})\to\IZ^n_2$ in \long\
is onto. (The partial section $\omega^{(1)}$ does not exist because
$t_1\in H^1(M_{NO};\pi_0(O(d)))$, which is the first Stiefel-Whitney class
$w_1$, is nontrivial.) Thus, unlike the orientable case, the group
$H=Im (p_n)_*$ is a proper subgroup of $B_n(M_{NO})$; it is the subgroup
generated by local exchanges of the identical particles, and single particle
excursions around {\it orientation preserving} loops in $M_{NO}$. However for
$n=1$ we can always reduce the calculation back to the orientable case by
noting that $F(M_{NO})$ is homeomorphic to $F({\tilde M})$, where ${\tilde M}$
is the {\it orientable double cover} of $M_{NO}$. Hence, $B^F_1(M_{NO})
[=\pi_1(F(M_{NO}))]$ is isomorphic to $B^F_1({\tilde M})
[=\pi_1(F({\tilde M}))]$. This equality persists to larger values of $n$ if
$d\geq 3$, as can be seen from the discussion following \ano . That is, we
have:
\vskip 4pt
\noindent
{\it If $M_{NO}$ is a nonorientable manifold of three or more dimensions,
and ${\tilde M}$ is its orientable double cover, then $B^F_n(M_{NO})=
B^F_n({\tilde M})$ for all $n\geq 1$.}
\vskip 4pt
\noindent
Thus, in this case, we may compute $B^F_n(M_{NO})$ using ${\tilde M}$ and the
methods described in sections 3 and 4. The interpretations of the generators
of $B_n^F({\tilde M})$ given in these sections remain valid in
$B_n^F(M_{NO})$. They now represent particle rotations, exchanges and loops on
$M_{NO}$. It is important to note that the above result is not generally valid
for $d=2$. For example, it can be shown that $B^F_2(\IR\IP^2)$ is not
isomorphic to $B^F_2(S^2)$.

For any closed, nonorientable surface $M^{(2)}_{NO}$ it has been shown that
$\pi_2(Q_n(M^{(2)}_{NO}))=\{ e\}$ (for $\IR\IP^2$ see \fad , for all higher
genuses see \bgt ). Thus, \ex\ becomes (compare to \yie )
\eqn\yiet{\{ e\}\rightarrow\IZ^n\buildrel i_*\over\rightarrow
B_n^F(M_{NO}^{(2)})\buildrel (p_n)_*\over\rightarrow H\rightarrow\{ e\},}
where $H$ is the subgroup of $B_n(M_{NO}^{(2)})$ described above. The ordinary
braid groups $B_n$ of closed, nonorientable surfaces are all known (for
$\IR\IP^2$ see \ref\vbt{J.~Van~Buskirk, Trans. Amer. Math. Soc. {\bf 122}
(1966) 81.}, for higher genuses see \ref\scot{G.~P.~Scott, Proc. Camb. Phil.
Soc. {\bf 68} (1970) 605.}), and in principle the subgroup $H$ can be
determined from these presentations. (For example, it can be shown from \vbt\
that $H\subset B_2(\IR\IP^2)$ is isomorphic to $\IZ_4$.) However, in all but
the simplest cases this is a formidable task. If $H$ can be determined the
next step is to compute the extension in \yiet , which then can be used to
discuss the spectrum of spin and statistics for identical particles on
$M^{(2)}_{NO}$. We leave the completion of this program for future
investigations.

\newsec{Relations to the $O(d+1)$-Invariant Sigma Model in
$(d+1)$-Dimensions}

There is a class of nonlinear field theories whose topological properties
are intimately related to those of the above quantum mechanical systems.
These are the $O(d+1)$-invariant sigma models in $(d+1)$-dimensions. At any
fixed time, such a system is described (classically) by a map from the space
manifold $M$ (which we assume to be closed and orientable) to the target space
$S^d$, $d$ = dim $M$. Thus, the configuration space is the set $X(M)\equiv
{\rm Map}(M,S^d)$ of all such maps (with the compact-open topology). This model
has topological solitons labelled by the {\it degree} $n$ of the map from $M$
to $S^d$ ($\pi_0(X(M))=\IZ$). Thus, we may write $X(M)$ as
$\cup_{n=-\infty}^{\infty}X_{n}(M)$, where the component $X_{n}(M)$
contains only the maps of degree $n$. A degree $n$ soliton configuration
$\phi_{n}:M\to S^d$ can be constructed by choosing a $d$-disk $D$ in $M$
and letting $\phi_{n}$ be constant everywhere except in the interior of
$D$. The disk $D$ with its boundary identified to a point can be thought of
as a $d$-sphere and $\phi_{n}$ is chosen to map this sphere $n$ times
around the target $S^d$. All other maps in $X_{n}$ are homotopic to this
configuration. In particular, $\phi_{n}$ is homotopic to a map consisting
of $n$ isolated solitons of degree one which resembles the $n$ identical
particle configurations on $M$ encountered in the quantum mechanical
examples discussed earlier. Indeed, for each $n\geq 1$ there is a continuous
map $f:Q_n^F(M)\to X_n(M)$ obtained by choosing $n$ nonoverlapping disks
centered at the positions of the framed particles in each configuration $q\in
Q_n^F(M)$, and then inserting a soliton of degree one on each of them. The
radius of these disks (in a fixed metric on $M$) can be taken as, say,
one-fourth of the distance between the two closest particles in $q$. The frame
on a given particle is used to determine the orientation of the corresponding
soliton. Since $f$ induces a homomorphism $f_*:\pi_1(Q_n^F(M))\to
\pi_1(X_n(M))$, each of the elements of the braid group $B_n^F(M)=
\pi_1(Q_n^F(M))$ has a counterpart in $\pi_1(X_n(M))$. We now compute the
groups $\pi_1(X_n(M))$, whose representations provide information about the
spectrum of spin and statistics for the above solitons.

We begin by reviewing some known results \ts\apb\ref\fink{D.~Finkelstein and
J.~Rubinstein, J. Math. Phys. {\bf 9} (1968) 1762\semi F.~Wilczek and A.~Zee,
Phys. Rev. Lett. {\bf 51} (1983) 2250.}. Corresponding to the
exchanges $\sigma_i$, the $2\pi$-rotations $r_j$ and the one-particle loops
$x^{(k)}$ in $B_{n}^F(M)$, there will be the analogous operations
$f_*(\sigma_i)$, $f_*(r_j)$ and $f_*(x^{(k)})$ for the $n$ degree one solitons.
These operations can be shown to generate $\pi_1(X_n(M))$. However, due to the
possibility of creation and annihilation of soliton-antisoliton pairs, the
$f_*(\sigma_i)$'s are all homotopic to each other. We denote their common
homotopy class by $\sigma$. A similar result holds for the $f_*(r_j)$'s and we
denote the associated class by $r$. Moreover, there is a {\it topological
spin-statistics relation} which states that $\sigma=r$. Finally, for $d\geq 3$
it can be shown that any two loops $f_*(x^{(k)})$ and $f_*(y^{(l)})$ commute,
even if $k=l$. We may thus obtain a partial presentation of $\pi_1(X_{n}(M))$
by taking the image under $f_*$ of the generators and relations for $B_n^F(M)$,
and adding the new relations\foot{For $n\geq 2$, the relations in (7.2) are a
consequence of those in $B^F_n(M)$ and (7.1). For $n=1$ (7.1) is vacuous, while
(7.2) in general provides new relations.}
\eqn\nr{f_*(\sigma_i)=f_*(r_j)\equiv r \qquad 1\leq i\leq n-1,\ 1\leq j\leq n}
and \eqn\nrt{f_*(x^{(k)})f_*(y^{(k)})=f_*(y^{(k)})f_*(x^{(k)})\qquad 1\leq
k\leq n,\ d\geq 3.} Although there could, in principle, be additional relations
that are needed, we will demonstrate that this is not the case; the above
procedure defines a complete presentation of $\pi_1(X_{n}(M))$. In other words,
if we denote by $G_n(M)$ the group obtained from $B_n^F(M)$ by adding the
relations in \nr\ and \nrt , then $\pi_1(X_n(M))=G_n(M)$.

Before going further, we wish to point out that although the map $f$ above is
only defined for $n$ positive, it is still possible to speak of the spin and
statistics (as well as other properties) of the degree one solitons in $X_n(M)$
for {\it any} $n\in\IZ$ --- even $X_{0}(M)$ which has
no analog in our mechanical systems. Degree one soliton rotation and
loop operations in an arbitrary $X_n(M)$ can be described (for example) by
starting with a configuration where the field is constant everywhere except on
a disk $D$ in $M$. Next, create a degree one soliton anti-soliton pair in the
vacuum outside of $D$, and then perform the appropriate operation on the
soliton before finally annihilating the pair. To define the soliton exchange
operation we must create {\it two} degree one soliton anti-soliton pairs,
exchange the two solitons, and then annihilate. The group $\pi_1(X_n(M))$ for
any $n$ is generated by (the homotopy classes of) these processes. (Indeed,
for $n\geq 1$ this is the same set of generators as that described above using
$f_*$.) We also note that there is a homomorphism $\psi_{n,m}:
\pi_1(X_n(M))\to\pi_1(X_m(M))$, for any $n,m\in\IZ$, which sends each such
degree one process in $X_n(M)$ to the corresponding one in $X_m(M)$. This
shows, among other things, that the spin-statistics connection holds in every
$X_n(M)$.

The starting point for our demonstration of the equality $\pi_1(X_n(M))=G_n(M)$
is the exact sequence \ref\llt{L.~L.~Larmore and E.~Thomas,
Math. Scand. {\bf 47} (1980) 232.}
\eqn\lt{H^{d-2}(M;\IZ)\buildrel \theta\over\to
H^d(M;\pi_{d+1}(S^d))\buildrel\lambda\over\to \pi_1(X_{n}(M))\buildrel
p\over\to H^{d-1}(M;\IZ)\to \{ e\} ,} which may be obtained using a
{\it Postnikov decomposition} of the target sphere $S^d$. For $d=2$ we have
$\pi_3(S^2)=\IZ$ and the map $\theta$ in \lt\ is given by $\theta
(x)=y^{2|n|}$, where $x\in H^0(M;\IZ)=\IZ$ and $y\in H^2(M;\IZ)=\IZ$ are
suitably chosen generators. For $d\geq 3$ we have $\pi_{d+1}(S^d)=\IZ_2$
and $\theta$ is equal to the composite ${\rm Sq}^2\circ\rho$, where
$\rho:H^{d-2}(M;\IZ)\to H^{d-2}(M;\IZ_2)$ is mod 2 reduction and
${\rm Sq}^2:H^{d-2}(M;\IZ_2)\to H^d(M;\IZ_2)$ is the {\it Steenrod square}
operation. Since $M$ is closed and orientable we may write (for any $d\geq
2$) $H^d(M;\pi_{d+1}(S^d))=\pi_{d+1}(S^d)$, and by Poincar\'e duality
$H^{d-1}(M;\IZ)=H_1(M)$. Thus, \lt\ becomes
\eqn\ltt{\{ e\}\to\pi_{d+1}(S^d)/Im\ \theta\buildrel\lambda\over\to
\pi_1(X_{n}(M))\buildrel p\over\to H_1(M)\to \{ e\} .}
The image under $\lambda$ of the generator of $\pi_{d+1}(S^d)$ is the
$2\pi$-rotation $r$ \ref\tpr{D.~C.~Ravenel and A.~Zee, Comm. Math. Phys.
{\bf 98} (1985) 239\semi A.~S.~Schwarz, Mod. Phys. Lett. {\bf A4} (1989) 403.}.
Under the homomorphism $p$, the operation of taking a degree one soliton
around a loop $\ell$ in $M$ goes to the homology class of $\ell$.

We now assume $d\geq 3$. Then, using the techniques described (for example) in
\ref\vlh{V.~L.~Hansen, Trans. Amer. Math. Soc. {\bf 265} (1981) 273.}, it can
be shown that each of the homomorphisms $\psi_{n,m}:\pi_1(X_n(M))\to
\pi_1(X_m(M))$ above is an isomorphism. (The isomorphism class of $G_n(M)$ is
similarly $n$-independent and we will simply write $\pi_1(X(M))$ and $G(M)$ in
what follows.) The exact sequence \ltt\ becomes
\eqn\lttt{\{ e\}\to\IZ_2/Im\ \theta\buildrel\lambda\over\to \pi_1(X(M))
\buildrel p\over\to H_1(M)\to \{ e\} ,}
where $\theta ={\rm Sq}^2\circ\rho$. It is also known that ${\rm Sq}^2(x)$ is
equal to the {\it cup product} $x\cup U_2$, where $x\in H^{d-2}(M;\IZ_2)$ and
the {\it Wu class} $U_2\in H^2(M;\IZ_2)$ is equal to $w_2+w_1\cup w_1$. Here
$w_1$ and $w_2$ are the first and second Stiefel-Whitney classes of $\tau_M$
respectively. Since $M$ is orientable we have $w_1=0$. Hence ${\rm Sq}^2(x)
=x\cup w_2$. We will now show that:
\vskip 4pt
\noindent
{\it The homomorphism $\theta$ is
trivial if and only if $M$ has no $H_2$-obstruction to a spin structure.}
\vskip 4pt
\noindent
To prove this we use Lemma 1 of \ref\mas{W.~S.~Massey, Proc. Amer. Math. Soc.
{\bf 13} (1962) 938.} which implies that
$\theta$ is trivial if and only if $w_2$ is the mod 2 reduction of an
integral class $z_2\in H^2(M;\IZ)$ which has finite order. Thus,
all we need to show is that there exists such an element $z_2$ if and only
if $M$ has no $H_2$-obstruction. We first tackle the ``only if''
implication. Manifolds with no $H_2$-obstruction fall into two classes.
First the spin manifolds which can be handled easily since they have
$w_2=0$. (Simply take $z_2=0$.) The remaining spaces are those with an
$H_1$-obstruction. If we write by the {\it Universal Coefficient Theorem}
\eqn\uct{\{ e\}\to {\rm Ext}(H_1(M),\IZ_2)\buildrel\alpha\over\to H^2(M;\IZ_2)
\buildrel\beta\over\to {\rm Hom}(H_2(M),\IZ_2)\to\{ e\} ,}
then the Stiefel-Whitney class $w_2$ for such a space lives completely
in the Ext subgroup. As such, it can be seen as the mod 2 reduction of an
element $z_2\in {\rm Ext}(H_1(M),\IZ)\subseteq H^2(M;\IZ)$ which clearly has
finite order. Finally, to deal with the ``if'' implication, we must show that
if $M$ {\it does} possess an $H_2$-obstruction, then no such element $z_2$ of
finite order exists. This follows from the fact that $\beta(w_2)\in
{\rm Hom}(H_2(M),\IZ_2)$ is the transgression $\tau$ (see Section 4), which is
nontrivial for these spaces. So if $w_2$ is the reduction of an integral class
$z_2$, then $z_2$ must map nontrivially into ${\rm Hom}(H_2(M),\IZ)$ which is
torsion-free. This completes the proof.

As a result of the above theorem, we see that for manifolds having no
$H_2$-obstruction to a spin structure we have
\eqn\ltsp{\{ e\}\to\IZ_2\buildrel\lambda\over\to \pi_1(X(M))\buildrel p\over\to
H_1(M)\to \{ e\} .}
We may write down a similar sequence for the group
$G(M)$ in this case. That is, it is straightforward to show (using the
results of Sections 3 and 4) that $G(M)/\IZ_2=H_1(M)$, where the normal
subgroup $\IZ_2$ is generated by the $2\pi$-rotation $r$. Since adjoining any
additional (nonredundant) relations to $G(M)$ would ruin this property, we
must have $\pi_1(X(M))=G(M)$ here. For spin manifolds $M_{spin}$, one can use
the results of Section 3 to further demonstrate that
$\pi_1(X(M_{spin}))=G(M_{spin})=\IZ_2\times H_1(M_{spin})$.  By contrast, the
extension \ltsp\ {\it does not split} if $M$ has an $H_1$-obstruction.
However $\pi_1(X(M))=G(M)$ is still abelian here --- for
instance, $\pi_1(X(\IR\IP^{4m+1}))=\IZ_4$ for $m\geq 1$. In the case
of manifolds with an $H_2$-obstruction, the map $\theta$ is onto. Hence
the rotation $r$ is trivial and from \lttt\ we have $\pi_1(X(M))=H_1(M)$ in
this situation. Using the results of Section 4 we see that $G(M)=H_1(M)$ as
well. We have thus shown that $\pi_1(X(M))=G(M)$ for all closed,
orientable manifolds of three or more dimensions. Note that unlike the case
of identical particles on $M$, there are only scalar quantizations of the
above nonlinear sigma models (for $d\geq 3$) since $\pi_1(X(M))$ is always
abelian. However, as in the particle case, there exists a one-dimensional IUR
of $\pi_1(X(M))$ yielding half-integral spin for the degree one solitons if
and only if $M$ has no $H_2$-obstruction.

For a closed, orientable surface $M^{(2)}_g$ of genus $g$, the extension
\ltt\ becomes \eqn\lts{\{ e\}\to\IZ_{2|n|}\buildrel\lambda\over\to
\pi_1(X_n(M^{(2)}_g))\buildrel p\over\to H_1(M^{(2)}_g)\to \{ e\} .} Two
major differences from the situation in higher dimensions are: (1)
$\pi_1(X_n(M^{(2)}_g))$ depends on $n$ (actually, only on $|n|$); and (2)
these groups are nonabelian for $g\geq 1$. Clearly we have
$\pi_1(X_n(S^2))=\IZ_{2|n|}$. The extension \lts\  for the groups
$\pi_1(X_n(M^{(2)}_{g\geq 1}))$ has been computed in \llt . They are
generated by degree one soliton loops $\rho_l$ and $\tau_l$,
$1\leq l\leq g$ (analogous to those defined in Section 5) along with the
degree one soliton rotation $r$. The defining relations are $r^{2n}=e$
along with
\eqn\sr{\eqalign{
\rho_lr&=r\rho_l\cr \tau_lr&=r\tau_l\cr
\rho_l\rho_k&=\rho_k\rho_l\cr \tau_l\tau_k&=\tau_k\tau_l\cr
\rho_l\tau_k&=\tau_k\rho_l\cr \rho_l\tau_l&=r^2\tau_l\rho_l}
\qquad\eqalign{&1\leq l\leq g,\cr
&1\leq l\leq g,\cr &1\leq l,k\leq g,\cr
&1\leq l,k\leq g,\cr &1\leq l\neq k\leq g,\cr
&1\leq l\leq g.\cr }}
Using the results of Section 5 it is easy to prove that
$\pi_1(X_n(M^{(2)}_g))=G_n(M^{(2)}_g)$ for $g\geq 0$ and $n\geq 1$ (see also
\jmr\apb ). Thus, the equality of $\pi_1(X_n(M))$ and $G_n(M)$ holds for all
closed, orientable manifolds of dimension two or more.

The above results can be extended to an (infinite volume) open manifold
${\hat M}=\IR^d\# M$ of dimension $d\geq 2$, whose one-point compactification
is the closed, orientable manifold $M$. Here the fields of the
$O(d+1)$-invariant sigma model must go to a constant at spatial infinity in
order to have finite energy. Thus, the space manifold ${\hat M}$ is effectively
compactified to $M$ and the appropriate configuration space is the set
$X^*(M)\equiv {\rm Map}_*(M,S^d)$ of all {\it basepoint preserving} maps from
$M$ to $S^d$. That is, for every $\phi_*\in X^*(M)$ we have $\phi_*(m_0)=s_0$,
where $m_0$ is the ``point at infinity'' on $M$ and $s_0$ is a fixed element of
$S^d$. The model still possesses topological solitons ($\pi_0(X^*(M))=\IZ$) and
we may again compute the group $\pi_1(X^*_n(M))$ for the degree $n$ sector
$X^*_n(M)$, $n\in\IZ$. There is a simple fibration relating $X^*(M)$ and the
space $X(M)$ considered previously. More precisely, for any $n$ there is a
fibering $\mu: X_n(M)\to S^d$ given by $\mu (\phi )=\phi (m_0)$ for all
$\phi\in X(M)$. The fiber above a given point $s_0\in S^d$ is just the
space $X^*_n(M)$. The long exact homotopy sequence of $\mu$ yields
\eqn\lehs{\cdots\to\pi_2(S^d)\to\pi_1(X^*_n(M))\to\pi_1(X_n(M))\to
\pi_1(S^d)\to\cdots\ \  .}
If $d\geq 3$, we see that $\pi_1(X^*_n(M))=\pi_1(X_n(M))$ and hence the
results of this section apply. (Note that $B_n^F({\hat
M})=B_n^F(M)$ in this case as well.) If $d=2$, all we know from \lehs\
is that $\pi_1(X_n(M^{(2)}_g))$ is a homomorphic image of
$\pi_1(X^*_n(M^{(2)}_g))$. These latter groups have been calculated in
\ref\gos{T.~R.~Govindarajan and R.~Shankar, Mod. Phys. Lett. {\bf A4} (1989)
1457\semi G.~Date, T.~R.~Govindarajan, P.~Sankaran and R.~Shankar, Comm. Math.
Phys. {\bf 132} (1990) 293.}; all one has to do is remove the relation
$r^{2n}=e$ from the above presentation of $\pi_1(X_n(M^{(2)}_g))$. In other
words $\pi_1(X^*_n(S^2))=\IZ$, while the groups
$\pi_1(X^*_n(M^{(2)}_{g\geq 1}))$ are generated by $r$, $\rho_l$ and $\tau_l$
subject only to the relations in \sr . Note that unlike the situation for
$X_n(M^{(2)}_g)$, these groups are independent of $n$ and (like the case of
particles on ${\hat M}_g^{(2)}$) the full range of fractional spin and
statistics can be obtained at the level of scalar quantizations. Finally, it
can be shown that $\pi_1(X^*(M^{(2)}_g))=G({\hat M}^{(2)}_g)$ for all
$g\geq 0$ \apb .

We conclude this section with a brief discussion of the $O(d+1)$-invariant
sigma model with a closed, {\it nonorientable} space manifold $M_{NO}$ of
dimension $d$. One major difference from the orientable case is that
$\pi_0(X(M_{NO}))=\IZ_2$. Thus there exist solitons, but they can be
continuously deformed into their antisolitons. (More precisely, solitons are
turned into antisolitons by bringing them around an orientation reversing loop
in $M_{NO}$.) Another way of saying this is that the degree of a map from
$M_{NO}$ to $S^d$ is only defined mod 2. Hence, we need only consider $X_0$ and
$X_1$. The exact sequence \lt\ is still valid for $M_{NO}$ (where $n$ is to be
taken mod 2), only now $H^d(M_{NO};\IZ)=\IZ_2$ for all $d$ and we cannot use
Poincar\'e duality on $H^{d-1}(M_{NO};\IZ)$. In place of \ltt\ we therefore
have
\eqn\tlt{\{ e\}\to\IZ_2/Im\ \theta\buildrel\lambda\over\to
\pi_1(X_{n}(M_{NO}))\buildrel p\over\to H^{d-1}(M_{NO};\IZ)\to \{ e\} ,}
where the map $\theta$ is still ${\rm Sq}^2\circ\rho$ for $d\geq 3$, but is now
trivial for $d=2$. The spin-statistics connection remains valid, and the
image under $\lambda$ of the generator of $\IZ_2$ is the homotopy class of the
2$\pi$-rotation of a degree one soliton. It has also been shown that
$\pi_1(X_n(M_{NO}))$ is abelian (and independent of $n$) for all $d$, and
further that \tlt\ splits for $d=2$ \llt . Note that there are two major
differences from the situation for identical particles on $M_{NO}$ (see Section
6). First, for any closed, nonorientable surface $M^{(2)}_{NO}$ the above
2$\pi$-rotation squares to the identity in $\pi_1(X_n(M^{(2)}_{NO}))
=\IZ_2\times H^1(M^{(2)}_{NO})$, while the analogous operation for particles
has infinite order (see \yiet ). Second, if $d\geq 3$ we have seen that
$B_n^F(M_{NO})=B_n^F({\tilde M})$, where ${\tilde M}$ is the orientable double
cover of $M$. However, in general, the groups $\pi_1(X_n(M_{NO}))$ and
$\pi_1(X_n({\tilde M}))$ are different\foot{One source of this difference is
the fact that $H^{d-1}(M_{NO};\IZ)\neq H_1({\tilde M})$ in general.} for any
$d$ --- for example, $\pi_1(X_n(S^1\times\IR\IP^2))\neq
\pi_1(X_n(S^1\times S^2))$. We leave details concerning the spin and statistics
of $O(d+1)$ solitons on nonorientable manifolds for future work.

\newsec{Half-Integral Spin on Nonspin Manifolds - A General Treatment}

The purpose of this section is to obtain a better understanding of why we
have encountered scalar quantum theories possessing half-integral spin objects
(in both mechanics and field theory) on manifolds having an $H_1$-obstruction
to a spin structure. We will also see how to get
similar results on certain spaces having an $H_2$-obstruction by
considering more general quantum theories than those based on IUR's of the
fundamental group of the relevant configuration space $Q$. Although we
restrict ourselves to scalar quantizations, we will consider state vectors
that are sections of an {\it arbitrary} complex line bundle $L$ over $Q$. Such
bundles can be classified by the elements of $H^2(Q;\IZ)$. By the Universal
Coefficient Theorem this group can be written as the direct product of the
free part of $H_2(Q)$ and the torsion part of $H_1(Q)$. The flat bundles
are associated with these latter torsion elements.\foot{Flat line
bundles associated with distinct one-dimensional representations of
$\pi_1(Q)$ may correspond to the same element of $H^2(Q;\IZ)$.} In
\ref\sor{R.~D.~Sorkin, Comm. Math. Phys. {\bf 115} (1988) 421.}, a
procedure is given which (for $d\geq 3$) allows us to determine the spin of
the above particle-like objects in the quantum theory associated
with a bundle $L$. First, we choose a configuration $q_0\in Q$ which
contains such an object and construct the map $g:SO(d)\to Q$ given by
$g(R)=Rq_0$. Here $R$ is a rotation in $SO(d)$ and $Rq_0$ denotes
the configuration in which the particle has been appropriately rotated. The
map $g$ induces a homomorphism $g^*:H^2(Q;\IZ)\to H^2(SO(d);\IZ)$. If the
line bundle $L$ corresponds to the element $z\in H^2(Q;\IZ)$, then the
particle has half-integral spin in the corresponding quantum theory if and
only if $g(z)$ is the nontrivial element of $H^2(SO(d);\IZ)=\IZ_2$. There is
always at least one integral spin quantization since we may choose $L$ to
be the trivial line bundle which corresponds to the identity element of
$H^2(Q;\IZ)$. If the particle-like objects live on a manifold $M$ (which for
the remainder of this section is assumed to be closed, orientable and of
dimension $d\geq 3$), then one may have guessed that there exist half-integral
spin quantizations if and only if $M$ is a spin manifold. However in previous
sections we have seen that this is not the case. More
precisely, such quantizations indeed exist if $M$
is a spin manifold, but they arise in certain other cases as well.
A study of these situations leads us to the following conjecture:
\vskip 4pt
\noindent
{\it In the above mechanical and field-theoretic models,
there exist half-integral spin scalar quantizations if and only if the
underlying space manifold $M$ possesses a} ${\rm spin}_c$ structure.
\vskip 4pt
\noindent
A manifold $M$ has a $spin_c$ {\it structure} if the principal
$SO(d)$-bundle $F(M)$ can be extended to a principal $Spin_c(d)$-bundle
over $M$, where $Spin_c(d)=Spin(d)\times U(1)/\IZ_2$. (The $\IZ_2$ action
sends $(S,u)\in Spin(d)\times U(1)$ to $(-S,-u)$.)
Such a structure exists on $M$ if and only if the second Stiefel-Whitney
class $w_2\in H^2(M;\IZ_2)$ is the mod 2 reduction of an element of
$H^2(M;\IZ)$. This is equivalent to the statement $b(w_2)=0$, where
${b:H^2(M;\IZ_2)\to H^3(M;\IZ)}$ is the {\it Bockstein homomorphism}.
Clearly all spin manifolds have a ${\rm spin}_c$ structure. Moreover, by the
discussion following \uct , all manifolds with an $H_1$-obstruction to a
spin structure possess ${\rm spin}_c$ structures. This is also true of some,
but not all, manifolds having an $H_2$-obstruction. For example the spaces
$\IC\IP^{2m}$, $m\geq 1$, are ${\rm spin}_c$ manifolds with an
$H_2$-obstruction. The same is true of the examples in \frm\ and \vas\ (see
\dui ). On the other hand, consider the five-dimensional coset
space $K=SU(3)/SO(3)$. We have $\pi_1(K)=\{ e\}$, $\pi_2(K)=\IZ_2$ and
$H^2(K;\IZ_2)=\IZ_2$. The generator of this last group can be shown to
be $w_2$. Thus, $K$ has a $\pi_2$- (and hence an $H_2$-) obstruction to
a spin structure. However because $H^2(K;\IZ)$ is trivial, $K$ is {\it not} a
${\rm spin}_c$ manifold \sstr . Since it is known that all closed, orientable
manifolds of dimension less than five are ${\rm spin}_c$, this example is in
some sense minimal. There are also examples of closed, flat Riemmanian
manifolds (in any even dimension $\geq 6$) which have a $\pi_1$- and an
$H_2$-obstruction to a spin structure and are not ${\rm spin}_c$ \dui .

We will first prove the above conjecture for the simple case of a single
particle on $M$. Here $Q=F(M)$ and
the map $g: SO(d)\to F(M)$ is just the inclusion map of an $SO(d)$ fiber into
$F(M)$. The induced map $g^*$ can be determined from the {\it Serre exact
cohomology sequence} for the frame bundle:
\eqn\ser{\{ e\}\to H^2(M;\IZ)\buildrel p^*\over\to H^2(F(M);\IZ)\buildrel
g^*\over\to H^2(SO(d);\IZ)\buildrel\beta\over\to H^3(M;\IZ)\ .}
It can be shown that $\beta$ maps the generator of $H^2(SO(d);\IZ)=\IZ_2$ to
$b(w_2)\in H^3(M;\IZ)$. Thus, $g^*$ is onto (showing the existence of a
half-integral spin scalar quantum theory) if and only if $M$ possesses a
${\rm spin}_c$ structure.\foot{Note that in order to construct half-integral
spin scalar quantum theories on ${\rm spin}_c$ manifolds having an
$H_2$-obstruction to an ordinary spin structure, such as $\IC\IP^{2m}$, we
must use nonflat bundles.} This result can be extended to the case of
$n$ {\it distinguishable} particles on $M$, $n\geq 1$. The configuration space
is ${\hat Q^F_n(M)}\equiv F^n(M)-\Delta_p$, and to find the spin of the $ith$
particle we use the map $g_i:SO(d)\to {\hat Q^F_n(M)}$ which rotates
this particle in a given configuration. We may also relate this
situation to the one-particle case through the commutative diagram
\eqn\com{\matrix{&SO(d) &\buildrel {\textstyle g_i}\over\longrightarrow
&{\hat Q}_n^F(M)\cr &\ &\searrow &\phantom{{\hat h_i}}\Big\downarrow {\hat h_i}
\cr &\ &{\displaystyle{g\phantom{}\atop\phantom{Y}}} &F(M),\cr }}
where the map ${\hat h_i}$ simply reads off the coordinates and frame of the
$ith$ particle. \com\ yields the following diagram of cohomology groups:
\eqn\tcom{\matrix{&H^2(SO(d);\IZ) &\buildrel {\textstyle g_i^*}\over
\longleftarrow &H^2({\hat Q}_n^F(M);\IZ)\cr &\ &\nwarrow
&\phantom{{\hat h_i^*}}\Big\uparrow {\hat h_i^*}\cr &\
&{\displaystyle{g^*\phantom{}\atop\phantom{Y}}} &H^2(F(M);\IZ).\cr }}
The commutativity of \tcom\ tells us that $g_i^*$ is onto if $M$ is
${\rm spin}_c$ (since $g^*$ is onto in this case). That is, for each $i$,
there exists at least one scalar quantum theory in which particle $i$ has
half-integral spin. Conversely, if $M$ is not a ${\rm spin}_c$ manifold then
$g^*$ is trivial. However this does not imply that $g_i^*$ is trivial, since
it is still possible that $g_i^*$ is onto and $Im\ {\hat h_i^*}\subseteq
Ker\ g_i^*$. To show that this does not occur, we will need to compute
$H^2({\hat Q_n^F(M)};\IZ)$. Without loss of generality we may assume that
$d\geq 5$, since all closed manifolds of dimension less than five are
${\rm spin}_c$. Since $\Delta_p$ has codimension $d$ in $F(M)^n$, the
inclusion of ${\hat Q_n^F(M)}$ into $F(M)^n$ induces an isomorphism on $H^2$
in this case.\foot{This is even true for $d=4$. For $d=3$, the induced map on
$H^2$ is an epimorphism.} So we can write $H^2({\hat Q_n^F(M)};\IZ)=
H^2(F(M)^n;\IZ)=H^2(F(M);\IZ)^n\times A$, where $A$ is free abelian and we have
used the {\it Kunneth formula} to obtain the last equality. (This result holds
whether or not $M$ is ${\rm spin}_c$.) The map ${\hat h_i^*}$
in \tcom\ is then an isomorphism onto the $ith$ factor of $H^2(F(M);\IZ)$ in
$H^2({\hat Q_n^F(M)};\IZ)$, and the commutativity of the diagram implies that
$g_i^*$ is trivial for all $i\geq 1$. Therefore, there are no half-integral
spin scalar quantizations for any particle. This completes the proof of our
conjecture for $n$ distinguishable particles on $M$.

At this point it would be natural to treat the case of $n$ {\it identical}
particles on $M$. However, we find it easier to first consider identical
$O(d+1)$ solitons on $M$ and then return to the particle situation. We
therefore wish to consider the groups $H^2(X_n(M);\IZ)$ for $M$ a closed,
orientable manifold of dimension $d\geq 3$. We will first treat the case $n=1$,
and then use this result to deal with all other values of $n$. We can relate
the $n=1$ case to our previous results for a single particle by using
the following commutative diagram:
\eqn\ri{\matrix{&SO(d) &\buildrel {\textstyle g}\over\longrightarrow
&F(M)\cr &\ &\searrow &\phantom{f}\Big\downarrow f \cr
&\ &{\displaystyle{j\phantom{}\atop\phantom{Y}}} &X_1(M).\cr }}
Here $j$ is the rotation map for a configuration containing a single degree one
soliton, and $f$ is the map of the previous section. This yields the
following diagram of cohomology groups:
\eqn\rit{\matrix{&H^2(SO(d);\IZ) &\buildrel {\textstyle g^*}
\over\longleftarrow &H^2(F(M);\IZ)\cr &\ &\nwarrow
&\phantom{f^*}\Big\uparrow f^* \cr &\
&{\displaystyle{j^*\phantom{}\atop\phantom{Y}}} &H^2(X_1(M);\IZ) .\cr }}
Without any explicit information about $f^*$, the commutativity of \rit\ along
with our results for the single particle case already show that there are no
half-integral spin scalar quantizations for degree one solitons in $X_1(M)$
when $M$ is not ${\rm spin}_c$. That is, in this case $g^*$ --- and hence
$j^*=g^*\circ f^*$ --- is trivial. To show that such quantizations exist
(or equivalently, $j^*$ is onto) if $M$ {\it is} ${\rm spin}_c$, we must
compute $H^2(X_1(M);\IZ)$. The torsion part of this group can be computed using
the results of the previous section on $\pi_1(X_1(M))$. The torsion-free part
can be found using standard techniques in the theory of localization of
homotopy types (in particular {\it rational homotopy theory})
\ref\loc{P.~Hilton, G.~Mislin and J.~Roitberg, {\it Localization of Nilpotent
Groups and Spaces} (North Holland, Amsterdam, 1975)\semi J.~M.~Moller and
M.~Raussen, Trans. Amer. Math. Soc. {\bf 292} (1985) 721.}. Putting these
results together, it is straightforward to show that $H^2(X_1(M);\IZ)$ has a
direct factor isomorphic to $H^2(F(M);\IZ)$. The map $f^*$ in \rit\ then sends
this factor onto the target $H^2(F(M);\IZ)$. The commutativity of \rit ,
along with the single particle result, then implies that $j^*$ is onto if $M$
is ${\rm spin}_c$. This takes care of the $n=1$ case. For an arbitrary $n$
consider a configuration in $X_n(M)$ which contains (among other things) an
isolated degree one soliton, and let $j:SO(d)\to X_n(M)$ be the rotation map
for this soliton. We must show that ${j^*:H^2(X_n(M);\IZ)\to H^2(SO(d);\IZ)}$
is onto if and only if $M$ is ${\rm spin}_c$. One can demonstrate (by an
extension of the arguments used to show the $n$-independence of
$\pi_1(X_n(M))$) that for any integers $n$ and $m$ there is an isomorphism
$\Psi_{n,m}:H^2(X_n(M);\IZ)\to H^2(X_m(M);\IZ)$ which makes
\eqn\rit{\matrix{&H^2(SO(d);\IZ) &\buildrel {\textstyle j^*}
\over\longleftarrow &H^2(X_m(M);\IZ)\cr &\ &\nwarrow
&\phantom{\Psi_{n,m}}\Big\uparrow \Psi_{n,m} \cr &\
&{\displaystyle{j^*\phantom{}\atop\phantom{Y}}} &H^2(X_n(M);\IZ) \cr }}
commute. By choosing $m=1$ and using our results for $X_1(M)$, we see that
for any $n\in\IZ$ the map $j^*$ is onto if and only if $M$ is ${\rm spin}_c$.
Thus, we have proven the conjecture for degree one solitons on $M$ in any
$X_n(M)$.\foot{We also note that the spin-statistics relation still holds in
the above general scalar quantizations of the $O(d+1)$-invariant sigma model
\sor .} Of course the same result holds for solitons of {\it any} odd degree,
while even degree solitons necessarily have integral spin.

Finally, we return to the case of $n$ identical particles on $M$. We will
use the commutative diagram ($n\geq 1$)
\eqn\ritt{\matrix{&\ &{\displaystyle{\phantom{X}\atop\phantom{} g_i}}
&{\hat Q_n^F(M)}\cr
&\ &\nearrow &\phantom{\chi_n}\Big\downarrow \chi_n \cr
&SO(d) &\buildrel {\textstyle {\bar g_i}}\over\longrightarrow &Q_n^F(M)\cr
&\ &\searrow &\phantom{f}\Big\downarrow f \cr
&\ &{\displaystyle{j_i\phantom{}\atop\phantom{Y}}} &X_n(M),\cr }}
where ${\bar g_i}$ (respectively, $j_i$) is the rotation map of the
particle (respectively, degree one soliton) in the $ith$ position in an $n$
identical particle (respectively, soliton) configuration, and $\chi_n$ is the
covering projection defined in Section 2. Again, we have the diagram of
cohomology groups
\eqn\crit{\matrix{&\ &{\displaystyle{\phantom{X}\atop\phantom{} g^*_i}}
&H^2({\hat Q_n^F(M)};\IZ)\cr
&\ &\swarrow &\phantom{\chi^*_n}\Big\uparrow \chi^*_n \cr
&H^2(SO(d);\IZ) &\buildrel {\textstyle {\bar g^*_i}}\over\longleftarrow
&H^2(Q_n^F(M);\IZ)\cr &\ &\nwarrow &\phantom{f^*}\Big\uparrow f^* \cr
&\ &{\displaystyle{j^*_i\phantom{}\atop\phantom{Y}}}
&H^2(X_n(M);\IZ).\cr }}
{}From the commutativity of the upper triangle in \crit\ and the result for
distinguishable particles, we see that there are no half-integral spin scalar
quantizations for the identical particles if $M$ is not a ${\rm spin}_c$
manifold. From the commutativity of the lower triangle and the result for
solitons we have that there exist half-integral spin scalar quantizations if
$M$ is a ${\rm spin}_c$ manifold. This completes the proof of the full
conjecture.

To summarize, we have considered the existence of half-integral spin scalar
quantum theories on a closed, orientable manifold $M$ of dimension $d\geq 3$.
We have shown that such quantizations exist for particles (identical or
not) or $O(d+1)$ solitons if and only if $M$ is a ${\rm spin}_c$
manifold.\foot{The result for $O(d+1)$ solitons can be extended to include
manifolds of the form $\IR^d\# M$, with $M$ closed and orientable, while that
for distinguishable particles is actually true for {\it any} orientable
manifold. Although we have no explicit proof, we believe that the result for
identical particles can also be extended to any orientable manifold.} Recall
that if $M$ has a $\pi_1$- and an $H_2$-obstruction to a spin structure, then
we have shown (see Section 4) that there exist half-integral spin
{\it nonscalar} quantizations for $n$ identical particles on $M$ {\it even if
$M$ is not $spin_c$} (as in the examples constructed in \dui ). Presumably
these quantizations utilize structures on $M$ more general than ${\rm spin}_c$
in order to obtain half-integral spin.

\newsec{Comments and Conclusions}

The purpose of this paper has been fourfold. First, to define a
generalization of the braid groups $B_n(M)$ which apply to identical
particles on $M$ having an internal structure described by a fiber bundle
$Y\hookrightarrow E\to M$, and to develop general techniques for the
computation of these new groups $B_n^E(M)$.\foot{For applications to particles
carrying an internal $U(1)$ charge in the presence of flux tubes, see
\ref\alice{L.~Brekke, W.~Fischler and T.~D.~Imbo, Phys. Rev. Lett. {\bf 67}
(1991) 3643\semi L.~Brekke and T.~D.~Imbo, UIC preprint UICHEP-TH/92-02.}.}
Second, to apply these techniques to the case of identical particles possessing
an internal spin degree of freedom --- that is, $E$ is the frame bundle $F(M)$
of the manifold $M$. Also, to construct the IUR's of the {\it spin braid
groups} $B_n^F(M)$ in order to discuss the spectrum of spin and statistics for
the particles. Third, to relate the groups $B_n^F(M)$ to the fundamental group
of the configuration space of the $O(d+1)$-invariant nonlinear sigma model with
space manifold $M$, $d$ = dim $M$. These models contain topological solitons
and the above relationship sheds light on their spin and statistics.
Finally, to discuss necessary and sufficient conditions on $M$ for the
existence of a (general) scalar quantum theory which yields half-integral
spin for either particles or solitons.

Our main results may be stated as follows. In Section 2 we displayed
$B_n^E(M)$ as an extension of a quotient group of $\pi_1(Y)^n$ by a subgroup
of $B_n(M)$. We also discussed the information about the bundle $E$ needed
to compute this extension, as well as some simplifications when $d\geq 3$.
In Sections 3 and 4 we were able to completely determine the groups
$B_n^F(M)$ for an arbitrary orientable manifold $M$ of dimension three or
more in terms of $\pi_1(M)$ and information about the possible obstructions
to a spin structure on $M$. More precisely, we showed that
$B_n^F(M)=(\IZ_2\times\pi_1(M))\wr S_n$ if $M$ is a spin manifold (Section
3), and $B_n^F(M)=B_n(M)=\pi_1(M)\wr S_n$ if $M$ has a $\pi_2$-obstruction
to a spin structure (Section 4). If $M$ has a $\pi_1$-obstruction then
$B_n^F(M)$ is an extension of $\IZ_2$ by $\pi_1(M)\wr S_n$
which is also described in Section 4. We gave a presentation of these
groups in terms of $2\pi$-rotations of the particle's frames, local
particle interchanges and single particle loops, which we used to obtain
information on the available spectrum of spin and statistics. In
particular, we were able to prove that there exists an IUR of $B_n^F(M)$
whose corresponding quantum theory  provides half-integral spin for the
identical particles if and only if $M$ does not have a $\pi_2$-obstruction
to a spin structure. Moreover, a {\it one-dimensional} IUR yielding
half-integral spin exists if and only if $M$ does not have an
$H_2$-obstruction. In Section 5 we discussed various
general properties of the groups $B_n^F(M)$ in two dimensions, and displayed
these groups when the orientable surface $M$ is closed. We further determined
$B_n^F(M)$ in terms of $B_n(M)$ when $M$ is flat. Section 6 contains a
brief discussion of nonorientable spaces. The main result is that the spin
braid groups of a nonorientable manifold and its orientable double cover
are isomorphic in three or more dimensions.

In Section 7 we turned our attention to the $O(d+1)$-invariant nonlinear
sigma model in $(d+1)$-dimensions with closed, orientable space manifold
$M$. We defined a homomorphism $f_*$ from $B_n^F(M)$ to the fundamental
group $\pi_1(X_n(M))$ of the degree $n$ sector $X_n(M)$ of the sigma model
configuration space, which sent particle exchanges, rotations and loops to
the corresponding operations for degree one solitons. We then demonstrated
that $f_*$ was onto for all $d\geq 2$. All one has to do is add the new
relations \nr\ and \nrt\ to our presentation of $B_n^F(M)$ in order to obtain
$\pi_1(X_n(M))$ --- each of these relations being a consequence of soliton
creation and annihilation processes. (In particular, there is a spin-statistics
relation for solitons.) For $d\geq 3$, the above result implies that the groups
$\pi_1(X_n(M))$ are always abelian, and hence all quantizations are scalar.
Moreover, as for particles, there is a scalar quantization yielding
half-integral spin for degree one solitons in $X_n(M)$ if and only if $M$ has
no $H_2$-obstruction to a spin structure. We closed the section with a
treatment of solitons on the open manifolds $\IR^d\# M$, proving results
analogous to the closed case, as well as a brief discussion of solitons on
nonorientable spaces. Finally, in Section 8 we set out to better understand
the half-integral spin quantizations we had encountered on certain nonspin
manifolds, as well as to show how to obtain such quantizations on other
nonspin manifolds, by considering quantum theories built from arbitrary complex
line bundles over the appropriate configuration space. We proved that there
exists such a general scalar quantum theory yielding half-integral spin for
particles or $O(d+1)$ solitons if and only if the closed, orientable space
manifold $M$ possesses a ${\rm spin}_c$ structure.

The results of this paper may be extended in several directions. For example,
in mechanics we may consider systems with $k$ distinct species of particles,
there being $n_i$ particles of species $i$. Working along similar lines we will
obtain the braid groups $B_{n_1,n_2,\dots ,n_k}^E(M)$ which are generalizations
of the ``partially colored'' braid groups $B_{n_1,n_2,\dots ,n_k}(M)$ in the
structureless case \pcb\gauge . We can further include the possibility of pair
creation and annihilation in these mechanical systems
\ts\apb\ref\crea{D.~McDuff,
Topology {\bf 14} (1975) 91\semi A.~P.~Balachandran, A.~Daughton, Z.-C.~Gu,
G.~Marmo, R.~D.~Sorkin and A.~M.~Srivastava, Mod. Phys. Lett. {\bf A5} (1990)
1575\semi A.~P.~Balachandran, R.~D.~Sorkin, W.~D.~McGlinn, L.~O'Raifeartaigh
and S.~Sen, Int. J. Mod. Phys. {\bf A7} (1992) 6887\semi A.~P.~Balachandran,
A.~Daughton, Z.-C.~Gu, R.~D.~Sorkin, G.~Marmo and A.~M.~Srivastava, Int. J.
Mod. Phys. {\bf A8} (1993) 2993.}. In field theory, we can consider much
broader classes of nonlinear sigma models. The existence and properties of
solitons in these systems will depend intricately on the nature of both the
space manifold and the target space (which will no longer be simply $S^d$).
For some general results and specific examples along this line, see
\diss\lbti\ltip\ref\sigm{L.~Brekke, S.~J.~Hughes and T.~D.~Imbo, UIC preprint
UICHEP-TH/92-18.}. (For instance, certain models with closed space manifolds
and nonsimply connected target spaces contain ambistatistical solitons
\lbti\ltip .) It would also
be interesting to further investigate the nature of half-integral spin
quantizations on {\it nonorientable} manifolds from the point of view of $pin$
and $pin_c$ structures \sstr , as well as to look at the possibility of
half-integral spin on ${\rm nonspin}_c$ manifolds by considering generalized
{\it nonscalar} quantum theories --- that is, quantum theories whose state
vectors are sections of an arbitrary complex {\it vector} bundle over the
appropriate configuration space. These latter theories may utilize structures
on the space manifold more general than ${\rm spin}_c$ (such as those in
\ref\avis{A.~Back, P.~G.~O.~Freund and M.~Forger, Phys. Lett. {\bf B77} (1978)
181\semi S.~J.~Avis and C.~J.~Isham, Comm. Math. Phys. {\bf 72} (1980) 103.})
in order to obtain half-integral spin. Finally, we note that a
discussion of the rather peculiar topic of particle spin and statistics on the
circle $S^1$ can be found in \ug\ref\cir{C.~Aneziris, A.~P.~Balachandran and
D.~Sen, Int. J. Mod. Phys. {\bf A6} (1991) 4721\semi C.~Aneziris, Int. J. Mod.
Phys. {\bf A6} (1991) 5047.}, while a reasonably complete treatment of
the spin and statistical properties of topological solitons on $S^1$ in a
sigma model with an arbitrary target space is given in \lbti .

\bigskip

\noindent
{\bf Acknowledgements }
\vskip 12pt

It is a pleasure to thank A.~K.~Bousfield, Cameron Gordon, Brayton Gray,
Anatoly Libgober, Ryan Rohm and Chandni Shah for helpful discussions.
This research was supported in part by the National Science Foundation under
grant PHY--92--18167, by the Texas National Research Laboratory Commission
under grants RGFY93-278 and RGFY93-278B, and by the U.~S.~Department
of Energy under contract numbers DE--AC02--89ER40509 and DE--FG02--91ER40676.
Most of this work was done while T.I. was a Junior Fellow in the Harvard
Society of Fellows.

\listrefs
\bye